\documentclass{aa}
\usepackage[dvips]{graphicx}
\usepackage{mathrsfs}
\usepackage{amsmath}
\usepackage{amssymb}
\usepackage[mathscr]{eucal}
\usepackage{latexsym}
\usepackage{amsbsy}
\usepackage{float}
\usepackage[latin1]{inputenc}
\newcommand{\bd}{\mathbf}
\def\BE{\begin{equation}}
\def\EE{\end{equation}}
\def\BA{\begin{eqnarray}}
\def\EA{\end{eqnarray}}
\def\BIT{\begin{itemize}}
\def\EIT{\end{itemize}}
\newcommand{\vt}{\vartheta}
\input epsf
\usepackage{natbib}
\bibpunct{(}{)}{;}{a}{}{,}

\def\be{\begin{equation}}
\def\ee{\end{equation}}
\def\ba{\begin{eqnarray}}
\def\ea{\end{eqnarray}}

\def\ltsima{$\; \buildrel < \over \sim \;$}
\def\lsim{\lower.5ex\hbox{\ltsima}}
\def\gtsima{$\; \buildrel > \over \sim \;$}
\def\gsim{\lower.5ex\hbox{\gtsima}}
\def\arcsecf {\hbox{$.\!\!^{\prime\prime}$}}

\begin{document}

   \title{Very weak lensing in the CFHTLS Wide: \\ Cosmology  from cosmic shear in the linear regime
\thanks{Based on observations obtained
 with {\sc MegaPrime/MegaCam}, a joint project of CFHT and CEA/DAPNIA, at
the Canada-France-Hawaii Telescope (CFHT) which is operated by the
National Research Council (NRC) of Canada,
the Institut National des Sciences de l'Univers of the Centre National de
la Recherche Scientifique (CNRS) of France, and the University of Hawaii.
This work is based in part
on data products produced at {\sc Terapix} and the Canadian Astronomy Data
 Centre as part of the Canada-France-Hawaii Telescope Legacy Survey, a collaborative project of
    NRC and CNRS.
}}

   \author{ L.~Fu$^{1,2}$, E.~Semboloni$^{1,3}$, H.~Hoekstra$^{4, }$\thanks{Alfred P. Sloan Fellow},
      M.~Kilbinger$^{1,3}$, L.~van Waerbeke$^{5}$, I.~Tereno$^{1,3}$,
      Y.~Mellier$^{1}$, C.~Heymans$^{1,5}$, J.~Coupon$^{1}$,
      K.~Benabed$^{1}$, J.~Benjamin$^{5}$, E.~Bertin$^{1}$,
      O.~Dor\'e$^{6}$, M.~J.~Hudson$^{7}$, O.~Ilbert$^{8,9}$,
      R.~Maoli$^{1,10}$, C.~Marmo$^{1}$, H.~J.~McCracken$^{1}$,
      B.~M\'enard$^{6}$} \offprints{fu@iap.fr}

  \institute{$^1$ Institut d'Astrophysique de Paris, UMR7095 CNRS,
   Universit\'e Pierre \& Marie Curie, 98 bis boulevard Arago, 75014 Paris,
   France \\
$^{2}$  Shanghai Normal University, 100 Guilin RD,  Shanghai, 200234, China \\
 $^3$ Argelander-Institut f\"{u}r Astronomie, Universit\"{a}t Bonn, Auf dem H\"{u}gel 71, 53121 Bonn, Germany \\
 $^4$ Department of Physics and Astronomy, University of Victoria, Victoria,
  B.C. V8P 5C2, Canada \\
 $^5$ University of British Columbia, Department of Physics and Astronomy, 6224
 Agricultural Road, Vancouver, B.C. V6T 1Z1, Canada \\
$^6$ Canadian Institute for Theoretical Astrophysics, University of Toronto
60 St. George Street Toronto, Ontario, M5S 3H8, Canada\\
 $^7$  Department of Physics and Astronomy, University of Waterloo, Waterloo ON N2L 3G1,
    Canada \\
$^8$ Laboratoire d'Astrophysique de Marseille, UMR 6110 CNRS-Université de  Provence, BP 8, 13376 Marseille Cedex 12, France\\
$^{9}$ Institute for Astronomy, 2680 Woodlawn Drive, Honolulu, HI
    96822-1897,  United States \\
 $^{10}$ Department of Physics, University La Sapienza, Pl. A. Moro 2, 00185,    Roma, Italy
}

\authorrunning{L.~Fu et al.}

\abstract%
{}%
{ We present an exploration of weak lensing by large-scale structure
in the linear regime, using the third-year (T0003) CFHTLS Wide data
release.  Our results place tight constraints on the scaling of the
amplitude of the matter power spectrum $\sigma_8$ with the matter
density $\Omega_{\rm m}$.  }%
{ Spanning 57 square degrees to $i^{\prime}_{AB} = 24.5$ over three
  independent fields, the unprecedented contiguous area of this survey
  permits high signal-to-noise measurements of two-point shear
  statistics from 1 arcmin to 4 degrees.  Understanding systematic
  errors in our analysis is vital in  interpreting  the
  results.  We therefore demonstrate the percent-level accuracy of our
  method using STEP simulations, an E/B-mode decomposition of the
  data, and the star-galaxy cross correlation function.  We also
  present a thorough analysis of the galaxy redshift distribution
  using redshift data from the CFHTLS T0003 Deep fields that probe
  the same spatial regions as the Wide fields.  }%
{ We find $\sigma_8(\Omega_{\rm m} / 0.25)^{0.64} = 0.785 \pm 0.043$
  using the aperture-mass statistic for the full range of angular
  scales for an assumed flat cosmology, in excellent agreement with
  WMAP3 constraints.  The largest physical scale probed by our
  analysis is 85 Mpc, assuming a mean redshift of lenses of 0.5 and a
  $\Lambda$CDM cosmology.  This allows for the first time to constrain
  cosmology using only cosmic shear measurements in the linear regime.
  Using only angular scales $\theta> 85$ arcmin, we find
  $\sigma_8(\Omega_{\rm m} / 0.25)_{{\rm lin}}^{0.53} = 0.837 \pm
  0.084$, which  agree with the results from our full
  analysis.  Combining our results with data from WMAP3, we find
  $\Omega_{\rm m}=0.248\pm 0.019$ and $\sigma_8=0.771\pm0.029$.}%
{}


\titlerunning{Very weak lensing in the CFHTLS Wide}

\maketitle

\section{Introduction}

A primary scientific goal of the Canada-France-Hawaii Telescope Legacy
Survey (CFHTLS\footnote{http://www.cfht.hawaii.edu/Science/CFHTLS/})
is the exploration of the properties of the dark matter power spectrum
and its evolution with redshift using weak gravitational lensing.  The
weak lensing signal manifests itself in a modification of the apparent
galaxy ellipticity induced by the cumulative weak gravitational shear
effects of large-scale structure (hereafter cosmic shear). The
statistical properties of the distortion field, as a function of
angular scale, reflect the properties of the Universe and of the dark
matter power spectrum projected along the line of sight \citep[see
reviews from][]{BS01,vWM03,Refregie03,Munshi06}.

The CFHTLS Deep and Wide surveys have been designed to maximise the
scientific reward of the CFHT {\sc MegaPrime/MegaCam} instrument and
in particular to produce a high-quality cosmic shear survey. The Deep
and Wide surveys provide image quality, depth and survey size
optimised for weak lensing studies as well as ($u^*,g',r',i',z'$)
colours over the whole field to get photometric redshifts \citep
{Ilbert06}.  Both depth and field of view have been increased by
roughly one order of magnitude as compared to the first-generation of
weak lensing surveys, like the Red Cluster Sequence \citep
[RCS,][]{2002ApJ...572...55H} and {\sc VIRMOS-Descart} \citep
{vWM00,vWMR01,vWM02,vWM05} surveys that were carried out at CFHT.

The first CFHTLS cosmic shear results were analysed by \cite {ESal06}
and \cite {HH06} who demonstrated that {\sc MegaPrime/MegaCam}
provides excellent quality data for weak lensing studies.  Despite the
optical distortion of the MegaPrime Wide field corrector, after
correction, no significant B-modes, nor any obvious critical
systematic residuals that may affect the weak lensing signal at the
percent level, were found.  The shear statistics as a function of
angular scale were in good agreement with the theoretical predictions
of the most popular cosmological models, and \cite {ESal06} showed
that the amplitude of shear signal convincingly increased with depth,
as expected from its sensitivity to redshift. These early CFHTLS
cosmic shear data were used by \citet {ESal06} and \cite {HH06} to
derive constraints on $\Omega_{\rm m}$-$\sigma_8$ and by \cite
{Schimd07} to explore some physical models of dark energy. The results
were consistent with the past CFHT weak lensing surveys but their
precision was still limited by the small sky coverage of the early
CFHTLS data and by the poor knowledge of the redshift distribution of
sources. \cite{Benjamin07}(hereafter B07) overcame these limitations
by using the early CFHTLS Wide data together with the Red Cluster
Sequence survey, {\sc VIRMOS-Descart} and the Garching-Bonn Deep
Survey \citep[GaBoDS,] [] {Hetterscheidt07} weak lensing surveys, and
the photometric redshifts of the joint CFHTLS-VIMOS VLT Deep Survey
(VVDS) analysis \citep {Ilbert06}. They then derived much more
reliable and accurate $\Omega_{\rm m}$-$\sigma_8$ constraint, lowering
the upper limits on $\sigma_8$ to be fully consistent with \cite
{Spergel07}.

The early CFHTLS weak lensing analysis, the joint B07 and the recent
Cosmic Evolution Survey studies \citep [COSMOS,][]{Massey07} explore
only small scale lensing. Their cosmological interpretation is
therefore sensitive to the non-linear evolution of the dark matter
power spectrum and several other physical and systematic effects that
primarily contaminate the lensing signal at small scales.  The most
serious are the high contribution of non-Gaussianity to the error
budget \citep {ESal07} and the signal contamination on scales below
$\sim$20 arcmin by the shear-shape correlation
\citep[][]{Hirata04,Heymans06} and by the intrinsic ellipticity
correlation \citep{KS02,Heymans03}.

The CFHTLS Wide survey has been designed to probe angular scales up to
8 degrees (the largest scale explored by all Wide fields). The
exploration of angular scales beyond one degree is technically
challenging due to the decreasing amplitude of the lensing signal.
Systematics in this unexplored territory are also still poorly
understood or unknown. However, they depend on the large-scale
accuracy and stability of field-to-field astrometric, photometric and
Point Spread Function (PSF) calibrations, and thus there is a need for
data homogeneity when analysing galaxy pairs separated by more than
the one-degree {\sc MegaCam} field of view.

The third release CFHTLS T0003 circumvents these issues: The T0003
Wide data explore angular scales up to 8 degrees, that is more than
one order of magnitude larger than the largest non-linear angular
scales.  It covers a total field of view slightly smaller than B07,
but with the great advantage of forming a single homogeneous sample
and of being easily calibrated using the CFHTLS-VVDS photometric
redshifts of \cite {Ilbert06} that are also derived from the T0003
release.

This work presents a weak lensing analysis of the CFHTLS T0003
$i'$-band Wide survey. It extends the previous analysis of the CFHTLS
Wide to angular scales up to 230 arc minutes (about 85 Mpc, assuming
$\Omega_{\rm m}=0.27$ and $h=0.72$, a flat Universe and a mean lens
redshift of 0.5). Its sky coverage is 57 square degrees, that is
nearly two times larger than early CFHTLS data and about 35\% of the
final CFHTLS wide sky coverage.  Furthermore, it includes a new
uncorrelated field, W2, providing a better estimate of the
field-to-field variance. The shear measurement pipeline is calibrated
and its performance is evaluated using simulated images produced by
the Shear TEsting Programme \cite[STEP,][]{step1,step2}.  The signal
error budget includes non-Gaussian corrections to the cosmic variance,
using the fitting formulae proposed by \cite {ESal07}. The effective
redshift distribution of sources is determined from the CFHTLS T0003
Deep survey and calibrated using the VVDS \citep {Ilbert06}.

The paper is organized as follows: In Sect.~2 we give a description of
the data set, including the image stacking procedure used in this
work.  In Sect.~3 we describe the production of weak lensing
catalogues.  After a brief review of the theoretical background, we
present the two-point shear results, together with the sky curvature
correction needed at large angular scales in Sect.~4.  The redshift
distribution is discussed in Sect.~5. In Sect.~6 we show the
cosmological parameter estimates, discuss the constraints from linear
scales and compare to other data sets. In Sect.~7 we discuss the
contamination to our weak lensing measurement from shear-shape
correlations.  Finally, we summarise and give our conclusions in
Sect.~8.

\section{Data Description}

\subsection{Overview of the CFHTLS T0003 release}

The Canada-France-Hawaii Telescope Legacy Survey (CFHTLS) is a 5-year
project set up jointly by the Canadian and French agencies.  The Deep
and Wide observations are all carried out in service mode by the CFHT
operation staff using the {\sc MegaPrime/MegaCam} instrument mounted
at the prime focus of the telescope.  The {\sc MegaCam} camera is
composed of an array of 9 $\times$ 4 CCDs (2048 $\times$ 4612 pixels
each). The pixel size at {\sc MegaPrime} focus is $0\arcsecf186$, so
that {\sc MegaCam} comprises a compact field of view of $1^\circ
\times 1^\circ$ \citep {Boulade03}.

Details on the Deep and Wide fields have been introduced in \cite
{ESal06} and \cite {HH06}, respectively.  After completion the W1, W2
and W3 Wide fields will be composed of 8$\times$9, 7$\times$7,
7$\times$7 different {\sc MegaCam} pointing positions,
respectively\footnote{\tt http://terapix.iap.fr/cplt/oldSite/Descart/
summarycfhtlswide.html}. Each centre position is separated by its
nearest neighbour fields by about one degree. For each field, a
sequence of 7$\times620$-second $i'$-band exposures, separated by a
small dither, is taken. The dithering pattern is encompassed within a
$3'\times4'$ box. Hence, neighboring pointings overlap in right
ascension by a minimum of two and a maximum of three arc minutes,
whereas the overlap in declination is bounded between three and four
arc minutes.  The overlapping regions are used for the
pointing-to-pointing internal astrometric calibration and
flux-rescaling processes.

The CFHTLS Wide T0003 release is produced from all {\sc MegaCam}
CFHTLS images obtained between June 1st, 2003 and September 5th, 2005,
that passed both the CFHT and initial Terapix validation
processes. Each individual raw image has been pre-processed
(bias/dark/fringe subtractions and flatfielding), CCD-to-CCD
flux-rescaled and photometrically calibrated at CFHT using the Elixir
pipeline \citep{Magnier2004}. The Elixir products are archived at CADC
and then transferred to the Terapix data centre for further higher
level processing and the production of the CFHTLS
releases\footnote{\tt Details on the T0003 release can be found at
http://terapix.iap.fr/rubrique.php?id\_rubrique=208}.

\begin{figure}[!b]
 \begin{center}
   \includegraphics[width = 9cm]{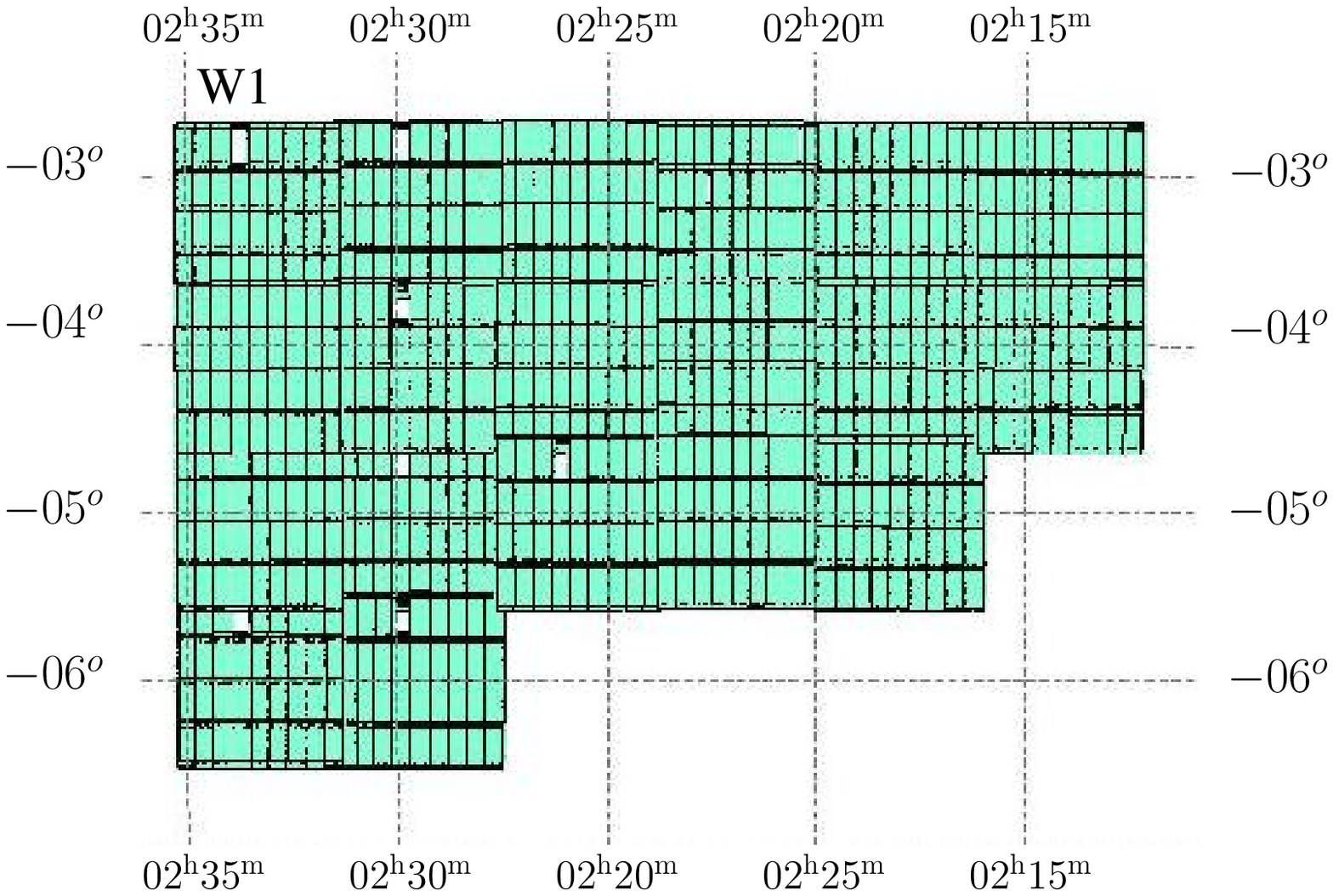}
   \includegraphics[width = 9cm]{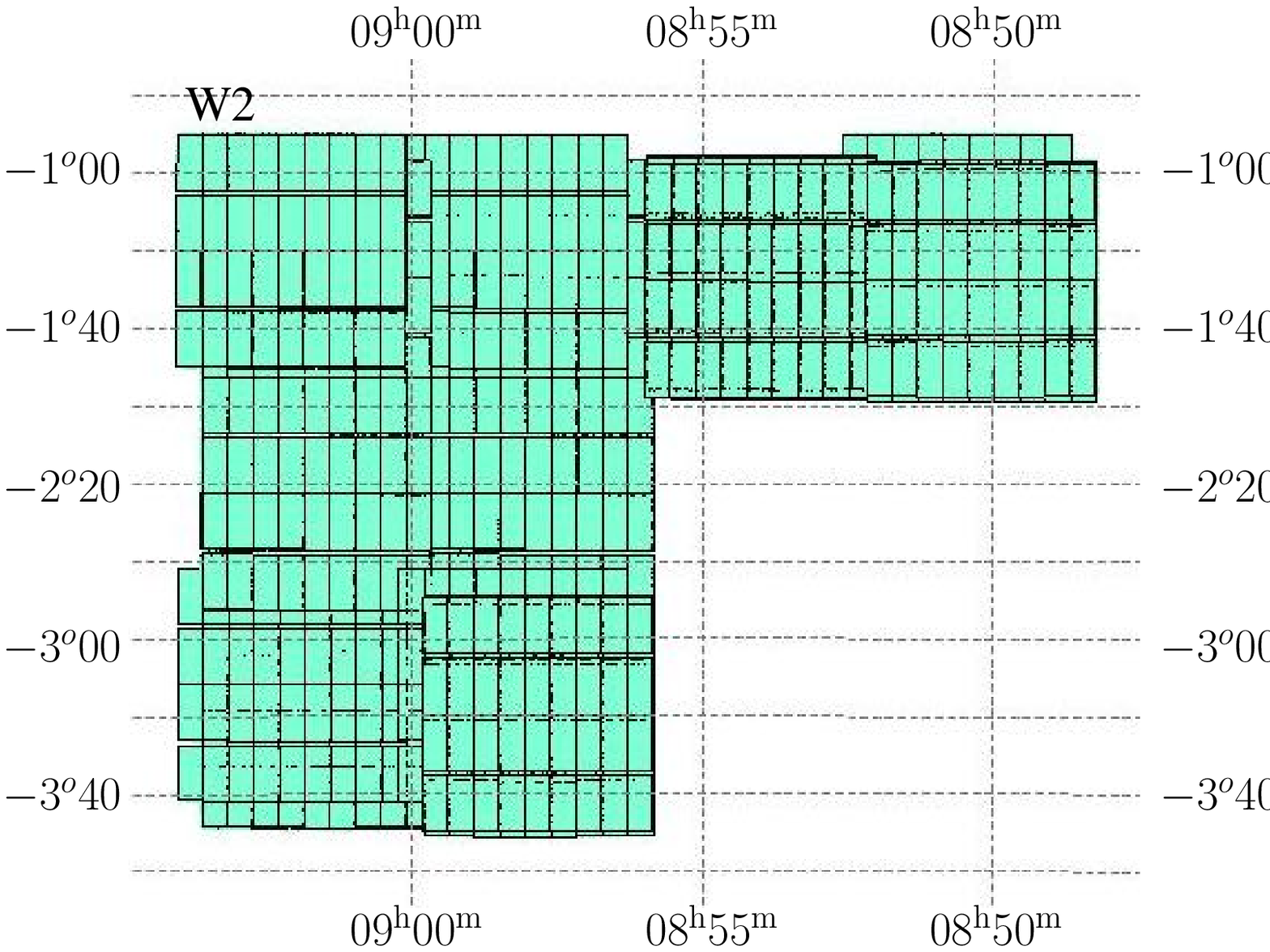}
   \includegraphics[width = 9cm]{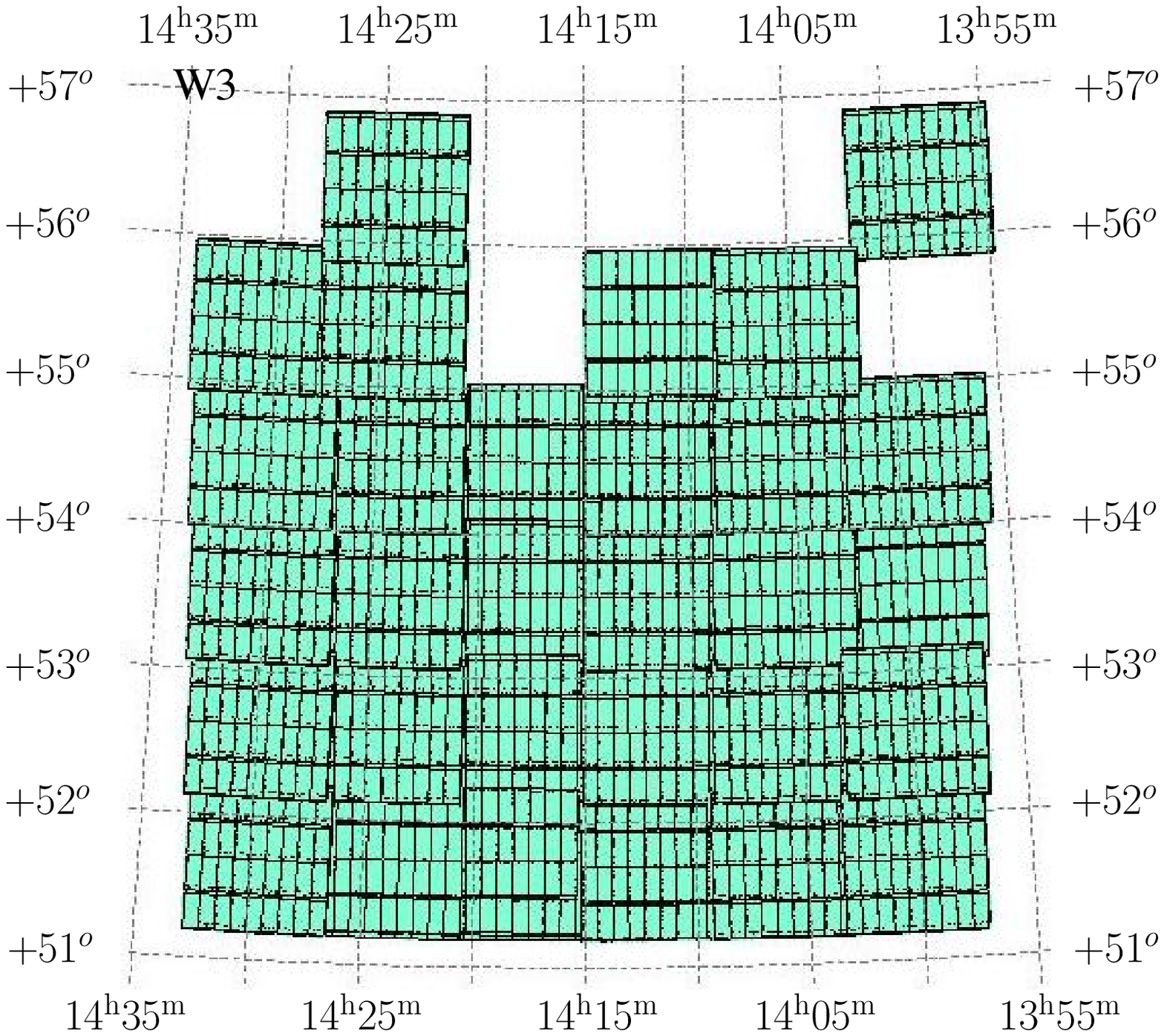}
   \caption{Sky coverage of the W1, W2 and W3 fields used in this
     work.  Each CCD is drawn as a small rectangle and each {\sc
     MegaCam} field is a squared mosaic of 36 rectangles. The small
     white holes are regions with missing data.  }
\label{sky_w}
\end{center}
\end{figure}

In this paper we use the $i'$-band pointings from the T0003 release,
but we introduce a more severe image selection in order to optimise
and homogenise the depth and the image quality over the whole field
used for weak lensing studies.  Each $i'$-band image must fulfil the
following second-level criteria before entering into the calibration
and stacking processes:

\begin{itemize}
\item half-light diameter of the Point Spread Function below $0\arcsecf9$;
\item individual exposure time larger than 500 seconds;
\item at least  4 exposures per stack.
\end{itemize}
Field-to-field seeing and depth variations are then minimised. Over
the 63 $i'$-band pointings, 57 pass the second level weak lensing
criteria and have been included in this work. They are distributed as
follows: 19 pointings on W1, 8 on W2, and 30 on W3 (see
Fig.~\ref{sky_w}). The W2 field is significantly less covered than W1
and W3 but is useful in order to derive an estimate of the CFHTLS Wide
field-to-field variance.

\subsection{Image Production}

The input stack images used in this work are produced using the same
Terapix procedures and software tools as the T0003 CFHTLS release.
Terapix first generates individual weight map images and individual
primary catalogues. It then proceeds to astrometric calibration, {\sc
MegaCam} field-to-field photometric rescaling, image re-centering,
image resampling, image warping and, finally, image stacking. Both a
co-added image and its weight map are produced, as well as an ASCII
DS9 readable mask file and a series of quality control meta-data.  A
description of these processing steps and software tools can be found
on the Terapix web site\footnote{{\tt http://terapix.iap.fr/soft}} as
well as on the Terapix release document \citep{Mellier05doc}. All
T0003 configuration files, parameter lists and processing command
lines are archived at CADC. Only $i'$-band images are considered in
this work since other filters cover a much smaller field of view, with
a large scatter in sky coverage and depth between each filter.

All fields are astrometrically calibrated and flux-rescaled using
SCAMP\footnote{{\tt http://terapix.iap.fr/soft/scamp}}
\citep{BERTIN05a,BERTIN06}. The astrometric reference catalogue is
USNO-B1, which is sufficiently accurate for the external astrometric
precision needed for this work. Internal astrometry and {\sc MegaCam}
pointing-to-pointing flux rescaling is done by identifying common
objects located in each overlap area. The image re-centering,
resampling and stacking are produced by SWarp\footnote{{\tt
http://terapix.iap.fr/soft/swarp}} \citep{BERTIN05b}, using the same
configuration and image processing parameters as those discussed by
\cite{HJMCC03}. All stacked images have a pixel size of
$0\arcsecf186$.

The astrometric calibration was performed for each pointing
individually. For each pointing, only exposures located inside a
circle of radius of 1.5 degrees were selected. This circle encompasses
all images at the centre field position together with all exposures
located around, at the 8 nearest neighbour centre positions. We
experienced that selecting images located beyond this radius did not
improve the accuracy and stability of the astrometric solution, and
sometimes would have even degraded it.  The 8 nearest neighbour fields
provide enough common stars in overlap regions to stabilise the
solutions at the boundary of each field. The internal $rms$ error
estimates of the astrometric calibration derived from the common
objects of nearest neighbour fields is $0\arcsecf030 \pm 0\arcsecf010$
for both {\sc MegaCam} directions, where the uncertainty denotes the
mean field-to-field scatter. The external $rms$ error is totally
dominated by the USNO-B1 internal error, which is $0\arcsecf35$ in
both directions.

In contrast, each stack does not use nearest neighbour images, but
only composes together a sequence of exposures having a small dither
with respect to a centre field.  Each stack is produced by SWarp,
using the weighted median value of each pixel and a Lanczos3
interpolation kernel. All output images have 19354$\times$19354 pixels
of $0\arcsecf186$, with North-East orientation along the $X$ and $Y$
pixel coordinates.  For all images we use a tangent projection and the
Equatorial J2000.0 astrometric coordinate system.

The reference photometric zero-point has been derived by CFHT using
standard Landoldt calibration fields (Landoldt 1992), but all
catalogues produced prior to weak lensing analysis have a default
zero~point magnitude set to 30.0. The magnitude system is instrumental
AB. An inspection of stellar colour-colour diagrams of each field
observed in 5 bands shows that the field-to-field scatter in the
overlapping regions is 0.03 mag. Comparison of SDSS and CFHTLS common
stars shows that the $i'$-band photometry agrees within 0.01-0.02
magnitude $rms$\footnote{\tt
http://terapix.iap.fr/article.php?id\_article=593}. However, only 10
W3 and 2 W1 fields have common objects with SDSS, so similar external
quality assessments cannot be done for all Wide pointings.

The mask files produced at image processing consist on a set of
polygons defined for each pointing in WCS coordinates. They mask the
periphery of each {\sc MegaCam} field of view and all halos and
saturated spikes produced by bright stars. In order to avoid
contamination by halos or diffusion from very luminous objects, all
bright stars located up to a radius of 45 arc minutes from the centre
position are automatically masked.  The size of polygons is scaled to
their apparent magnitude provided in the USNO-B1 catalogue.

The masks are then tuned by adding or modifying polygons from a visual
inspection of each stacked image.  This step is necessary to clean all
images from non-stellar contamination or stellar defects that were
missed by the automatic masking process. This includes big halos
produced by extremely bright stars, nearby galaxies or any features
that may produce a diffuse light component with sufficiently steep
gradient to contaminate the second moments of a galaxy's surface
brightness profile which is used to derive its ellipticity. Regions
with low signal-to-noise ratio are also masked. In particular, the
imprints of gaps between CCDs as well as the boundary of each field
are discarded and masked systematically.  They are revealed by
low-noise strips with a typical rectangular shape that draws the
border of each detector.  Finally, meteorite, asteroid and aeroplane
tracks that may still remain in the stacks are masked as well.  The
size of each polygon is generally significantly larger than the visual
size of the defect it masks. Using this conservative masking process,
the final effective sky coverage of the 57 selected Wide fields drops
to 34.2 deg$^2$, roughly 60\% of the total field.

\section{Production of weak lensing catalogues}

 Our shear measurement pipeline was optimised and calibrated using the
STEP1 and STEP2 simulations from \cite{step1} and \cite{step2}.  See
appendix A for a description of both our pipeline and the STEP
simulations. Table~\ref{pipe} lists the optimised parameter values of
our pipeline.

\subsection{Object selection}
\label{sec:objsel}

The lensing catalogue is generated by the {\tt IMCAT} software
\citep{KSB95}.  The size of each object is defined by the aperture
radius parameter $r_g$ given by the {\tt IMCAT} peak finding
algorithm.  The significance detection threshold, as defined by the
{\tt IMCAT} parameter $\nu$, is set to $\nu=8$ ({\it i.e.} above the
rms noise). This value was set according to the STEP tests in order to
maximise the number of objects detected while still keeping the bias
on the shear components negligible. The catalogue is then filtered to
remove objects with radius smaller than the seeing or larger than 6.75
pixels (about 1.3 arc seconds).  Pairs with angular separation smaller
than 10 pixels ($1\arcsecf86$) are also discarded in order to avoid
contamination from overlapping isophotes.

The magnitude of each object is derived by computing its flux within
an aperture radius of $3\times r_g$.  Only objects with {\tt IMCAT}
magnitude $21.5 \leq i'_{AB} \leq 24.5$ are kept into the final
analysis catalogue. Beyond this limit, the sample completeness drops
significantly below 50\%, most objects are too noisy and their shapes
are no longer reliable for the precision needed for weak lensing
studies.  The final catalogue based on the T0003 release of CFHTLS W1,
W2 and W3 fields contains roughly two million galaxies. Due to the
different weighting applied during the sample selection, the effective
number of galaxies used for the weak lensing analysis is 1.7 million,
spread over the effective area of 34.2 deg$^2$.  It corresponds to a
galaxy number density of 13.3 gal/arcmin$^2$.  The shapes of these
galaxies are quantified by measuring their ellipticities.

\subsection{PSF  correction}

\begin{figure}
\begin{center}
\includegraphics[width=7.5cm]{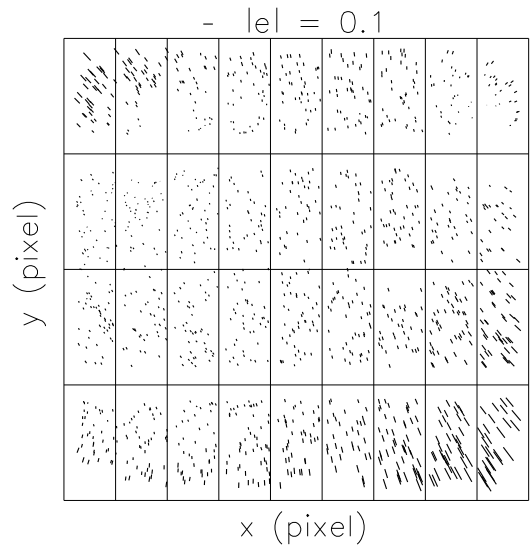}
\caption{The pattern of the PSF anisotropy in an example pointing
 ${\rm W}3+2+0-{\rm CFHTLS\_W}\_i\_143023+543031$.  Ticks represent
 the observed ellipticities at stellar locations.  On top of the
 figure a 10\% ellipticity modulus is shown for comparison.  }
\label{psf1}
\end{center}
\end{figure}

The ellipticities of galaxies are corrected from the PSF produced by
telescope, detector, optical and atmospheric effects, using the
Kaiser, Squires and Broadhurst method \citep {KSB95,LK97,HOEK98},
hereafter KSB+.  Our implementation of KSB+ is based on the one used
in \citet[] [ referenced as `LV' in Heymans et al.~2006a]
{vWM00,vWMR01,vWM02,vWM05}.  We calibrated it and modified its input
parameters after a new series of optimisations made with the STEP
simulations.  The results are presented in Appendix A. The version
used in this work recovers shear with an underestimation of only 1\%
to 3\% on the simulated images.

The PSF is measured at the stellar positions. After identifying the
stars in the images and assuming the PSF changes smoothly across the
field, the KSB-quantities known at the stellar positions can be
estimated at the galaxies positions by using a polynomial fit. The
typical pattern of the PSF anisotropy across one $1\deg^2$ field shows
a significant variation across the whole camera (Fig.~\ref{psf1})
which suggests the need to perform the fit in each CCD
separately. Each CCD covers $\rm{7 \times 4 \,\rm arcmin^2}$ and
contains an average of 43 stars, which allows an accurate mapping of
the PSF with a second order polynomial function.

A weight, $w$, is assigned to the ellipticity components of each
galaxy and used in the shear measurement (see Eq.~\ref{xiest}). We use
the \cite{HH02} weighting scheme
\begin{equation}
w=\frac{P_\gamma^2}{\sigma_\epsilon^2P_\gamma^2+\sigma_e^2},\;\,
\label{weightelip}
\end{equation}
where $\sigma_e$ is the error on the ellipticity measurement defined
in \cite{HH02} and $P_\gamma$ is a shear polarisability
\citep{LK97}. The weight also depends on the intrinsic ellipticity
dispersion, $\sigma_\epsilon$, which is derived from the mean
intrinsic ellipticity dispersion of the whole galaxy catalogue. We
find a value of $\sigma_\epsilon=0.42$.

\begin{figure}[!bt]
\begin{center}\includegraphics[width=7.5cm,height=7cm]{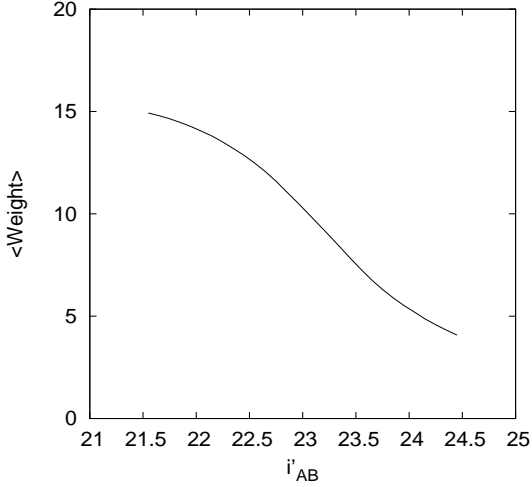}
\caption{The average galaxy weight (with arbitrary normalisation) as a
function of $i'_{AB}$ in the range of [21.5;24.5].} \label{weight}
\end{center}
\end{figure}

The shape of the weighting as function of the magnitude is shown in
 Fig.~\ref{weight}.  It decreases for fainter magnitudes since the
 error on the ellipticity increases when the signal-to-noise ratio
 decreases.

\section{Two-point cosmic shear statistics}


\subsection{Theoretical background}
\label{sec:theo}

The cosmic shear power spectrum is identical to the lensing
convergence power spectrum, $P_\kappa$, which is a projection of the
dark matter power spectrum, $P_{\rm \delta}$, along the line of sight
\citep[see for example][]{BS01}:
\begin{eqnarray}
P_\kappa(\ell)&=&{9\over 4}\Omega_{\rm m}^2
\left(\frac{H_0}{c}\right)^4\int_0^{\chi_{\rm lim}} {{\rm d}\chi \over
  a^2(\chi)}
P_{\delta}\left({\ell\over f_K(\chi)};
\chi\right)\nonumber\\
&&\times\left[ \int_\chi^{\chi_{\rm lim}}{\rm d} \chi' n(\chi') {f_K(\chi'-\chi)\over f_K(\chi')}\right]^2,
\label{pofkappa}
\end{eqnarray}
\\ where $\chi$ is the comoving distance along the light ray and
$\chi_{\rm lim}$ is the limiting comoving distance of the survey;
$f_K(\chi)$ is the comoving angular diameter distance; $n(\chi)$ is
the redshift distribution of the sources and $\ell$ is the modulus of
a two-dimensional wave vector perpendicular to the line of sight.
Equation~(\ref{pofkappa}) shows that the cosmological information contained
in the lensing power spectrum is degenerate with the redshift of the
sources.

The convergence power spectrum can be derived from the two-point shear
correlation functions. In particular, the $\xi_\pm$ correlation
functions relate to the power spectrum according to
\begin{eqnarray}
\xi_\pm(\theta) \equiv \xi_{\rm tt}(\theta) \pm \xi_{\times\times}(\theta)
=  {1\over 2\pi} \int_0^\infty {\rm d}\ell\,  \ell \, P_\kappa(\ell) {\rm J}_{0,4}(\ell\theta) \ , \ &&
\label{theogg}
\end{eqnarray}
where $\xi_{\rm tt}(\theta) $ and $\xi_{\times\times}(\theta)$ are the
tangential and rotated ellipticity correlation functions (given in
Eq.\ref{xiest}), $\theta$ is the angular separation between galaxy
pairs, and ${\rm J}_{0,4}$ are Bessel functions of the first kind.

Other two-point functions of the shear field may be derived from
$\xi_{\pm}$, such as the top-hat filtered variance of the shear and
the variance of the aperture-mass, in circular apertures
\citep{S02}. Respectively,
\begin{eqnarray}
\lefteqn{\langle |\gamma|^2\rangle _{\rm E, B}(\theta) = }
\nonumber \\ && \int_0^\infty
\frac{{\rm d} \vartheta \, \vartheta}{2 \theta^2} \left[ S_+\left(
\frac \vartheta \theta \right) \xi_+(\vartheta)
\ \pm \ S_-\left(\frac \vartheta \theta \right) \xi_-(\vartheta)
\right];
\label{tophateb}
\end{eqnarray}
and
\begin{eqnarray}
\lefteqn{\langle M_{\rm ap, \perp}^2\rangle(\theta)= }
\nonumber \\ & & \int_0^{2\theta}
\frac{{\rm d}\vartheta \, \vartheta}{2 \theta^2}
\left[ T_+\left(\frac\vartheta\theta\right)\xi_+(\vartheta) \ \pm \
T_-\left(\frac\vartheta\theta\right)\xi_-(\vartheta)\right] \ .
\label{mapeb}
\end{eqnarray}
 The filter functions $S_{+/-}$ and $T_{+/-}$ are defined in \cite{S02},
\BA
S_+(x)& =& \frac{1}{\pi} \left [4 \arccos(x/2)  - x\sqrt{4-x^2}\right
] {\rm H}(2-x) \ ;  \nonumber\\
S_-(x) &=& \frac{1} {\pi x^4} \times \nonumber\\
&& \left[ { x\sqrt{4-x^2}(6-x^2)-8(3-x^2) \arcsin(x/2)} \right ]\!\!,
\label{S}
\EA
\BA
T_+(x) & = & \frac{6(2-15x^2)}{5}\left[1-\frac 2 \pi
  \arcsin(x/2)\right] + \frac{x \sqrt{4-x^2}}{100 \pi} \nonumber  \\
   \times && \!\!\!\!\!\!\!\! \left(120+2320 x^2 -
754x^4+132x^6-9x^8 \right) {\rm H}(2-x)\ ;  \nonumber\\
T_-(x)  & = &  \frac{192}{35 \pi} x^3
\left(1-\frac{x^2}{4}\right)^{7/2} {\rm H}(2-x) \ ,
\label{T}
\EA
where $\rm H$ denotes the Heaviside step function.

All second-order statistics are different filtered versions of the
convergence power spectrum. Therefore they probe different properties
of the same power spectrum.

The cosmological shear field is (to first order) curl-free and is
called an E-type field. It is useful to decompose the observed shear
signal into E (non-rotational) and B (rotational) components. A
detection of non-zero B-modes indicates a non-gravitational
contribution to the shear field, which reveals a likely systematic
contamination to the lensing signal.  {\cite {Crittenden02} and \cite
{P02al}} derived an analogous decomposition for the shear
correlations, which is also used in this work:
\begin{equation}
\xi_{\rm {E, B}}(\theta)=\frac{\xi_+(\theta) \ \pm\
\xi'(\theta)}{2},
 \label{xieb}
\end{equation}
 where the definition of $\xi^\prime$ is also given in \cite{S02}
\BE
\xi^\prime(\theta) = \xi_-(\theta) + \int_\theta^\infty
\frac{\rm d\vt}{\vt} \xi_-(\theta)\left ( 4 - 12 \left( \frac{\theta}{\vt} \right)^2 \right ) \ .
\EE

Both $S_-$ and $\xi^\prime$ have infinite support, which implies the
E/B decomposition of the shear correlation function and of the top-hat
shear variance are non-local. They can be computed from data, up to an
offset value which depends on the largest angular separation
$\vartheta_{\rm max}$. This offset is a constant for $\xi_{\rm E,B}$
and a function of $\theta$ for $\langle |\gamma|^2\rangle _{\rm E,
B}(\theta)$.  In contrast, the aperture-mass variance decomposition is
local, providing an unambiguous decomposition. In practice, however, a
lower limit on the angular separation of galaxy pairs imposed by
contamination of overlapping isophotes may bias its amplitude.  The
lack of galaxy pairs closer than around $3\arcsec$ causes an
underestimation of the aperture-mass dispersion. This bias is small,
of order 5\% for $\theta=1'$ and smaller than one percent on scales
larger than 2 arc minutes \citep{K06}.

\subsection{Sky curvature correction at large angular scales}

The shear correlations are computed as follows:
\begin{equation}
\xi_{\rm tt}(\theta)=\frac{\sum w_i w_j
e_{{\rm t}}({\bf x}_i) \cdot e_{\rm t}({{\bf x}_j})}
{\sum w_i w_j}; \nonumber
\end{equation}
\begin{equation}
\xi_{\rm \times\times}(\theta)=\frac{\sum w_i w_j
e_{{\rm \times}}({\bf x}_i) \cdot e_{\rm \times}({{\bf x}_j})}
{\sum w_i w_j} \ ,
\label{xiest}
\end{equation}
where $\theta = |x_i-x_j|$ is the separation of pairs.  The
ellipticities are locally decomposed in each pair frame in a
tangential and a cross-component. The tangential component is computed
orthogonal to the line connecting each galaxy pair. The
cross-component is derived at a $\pi/4$ angle to the connecting
line. Using Eqs.~(\ref{tophateb})-(\ref{xieb}) we estimate the E and B
modes of the top-hat variance, of the aperture-mass variance and of
the shear correlations.

We correlate galaxies which are up to more than seven degrees
apart. At such large angles the curvature of the sky is no longer
negligible.  To avoid a potential bias due to projections we calculate
distances and angles in spherical co-ordinates as follows.

The distance $d$ between two objects at right ascension and
declination $(\alpha_i, \delta_i), i=1,2$, computed along the
great-circle connecting the 2 objects is given by
\begin{equation}
  \cos d = \cos(\alpha_1 - \alpha_2) \cos
  \delta_1 \cos \delta_2 + \sin \delta_1 \sin \delta_2 .
  \label{d_radec}
\end{equation}
 In order to decompose the ellipticity in tangential and cross
components, we need to consider the angle between the great circle
defined by the two galaxies and a parallel, since each pointing is
aligned with lines of constant declination. This is the course angle,
given by
\begin{equation}
  \tan \varphi = \frac{\sin(\alpha_1 - \alpha_2) \cos \delta_2}
       {\cos \delta_1 \sin \delta_2 - \sin \delta_1 \cos \delta_2
         \cos(\alpha_1 - \alpha_2)} .
       \label{course-angle}
\end{equation}

Not only the distances but also the galaxy ellipticities are affected
by the sky curvature. The ellipticity modulus of a galaxy remains
virtually unchanged, curvature on scales of a few arc seconds is
negligible. However, since ellipticity is also characterized by its
orientation, one has to be careful when correlating ellipticities at
large angular distances \citep{Castro2005}. In our case, the
ellipticity components $e_1$ and $e_2$ are measured in the local
Cartesian co-ordinate frame given by the $X$- and $Y$-axes of the
corresponding individual CFHTLS Wide pointing. Since each pointing is
projected using its own tangent point and defines its own co-ordinate
frame, we correlate ellipticities of galaxies from different pointings
by using their components measured in the respective local co-ordinate
systems.  In doing that, the sky curvature is neglected over the scale
of a single pointing but it is taken into account between pointings.
The effect in a single pointing corresponds to a small ellipticity
rotation for galaxies which are at a finite distance from the pointing
centre, at most 43 arc minutes.

 We compared the shear statistics computed using spherical
co-ordinates and using the following simple projection: Cartesian
co-ordinates ($X$, $Y$) of a galaxy with right ascension $\alpha$ and
declination $\delta$ are defined by $X=\alpha \cos \delta_{\rm c}$ and
$Y = \delta$, where $\delta_{\rm c}$ is the declination of the field
centre. The relative error is on the order of a couple of percent on
average. On larger scales, where the shear signal is small, this
relative error can be much higher. Therefore, throughout this paper we
take the sky curvature into account by calculating the shear
statistics in spherical co-ordinates.

\subsection{Results}

\begin{figure}
  \begin{center}
    \includegraphics[width=7cm,height=8cm,angle=-90, bb= 50 115 554 680]{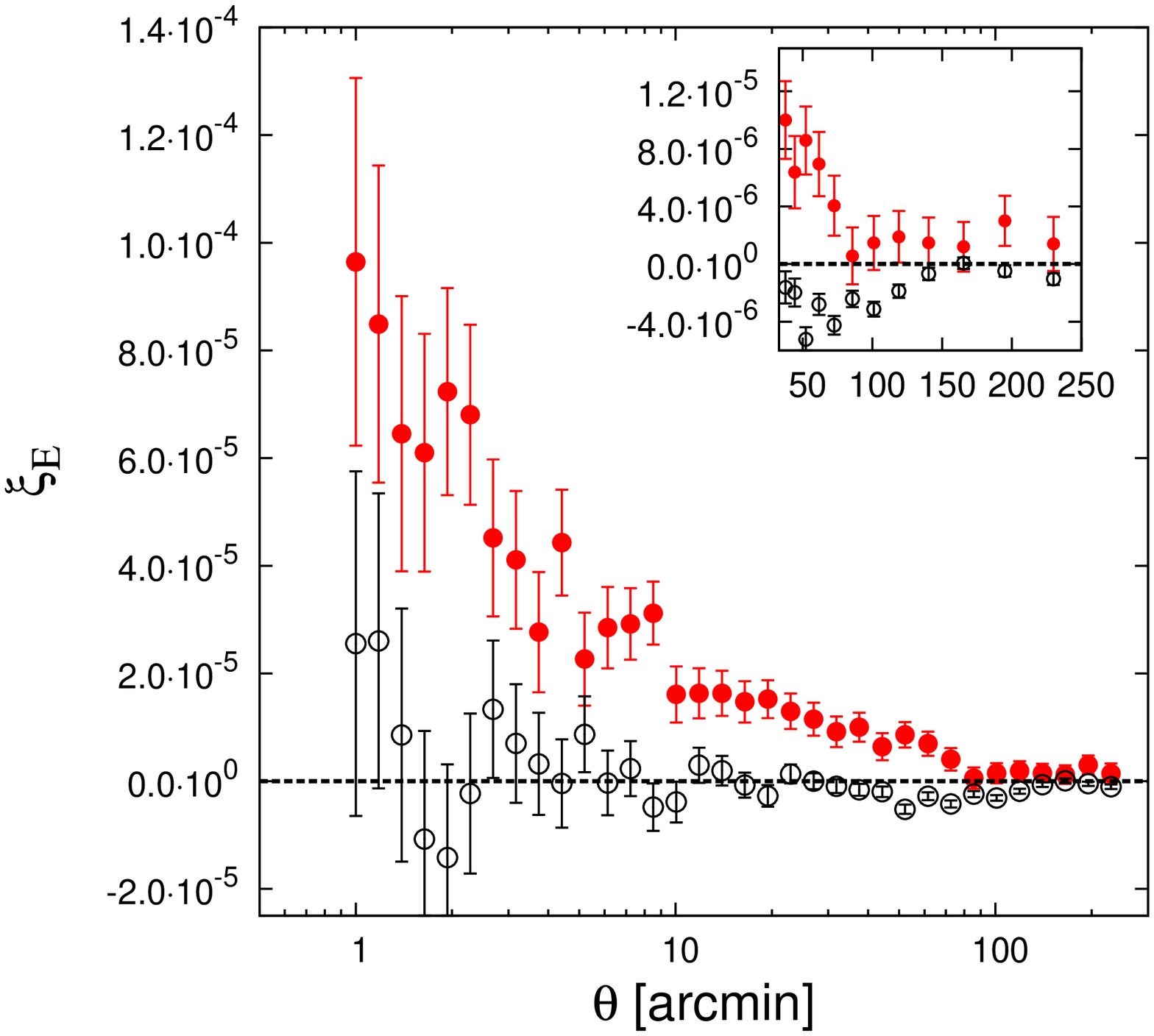}
    \includegraphics[width=7cm,height=8cm,angle=-90, bb= 50 115 554 680]{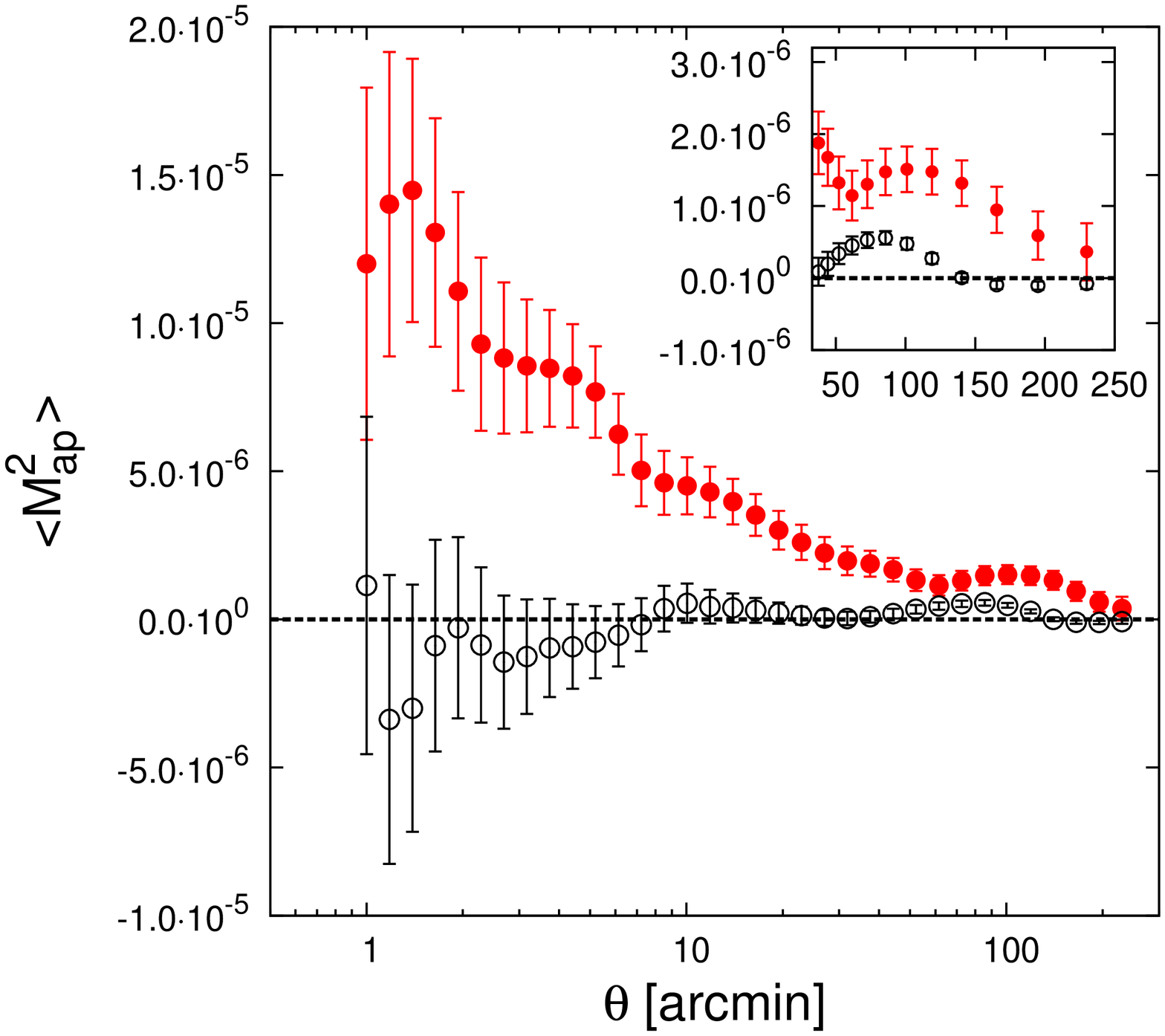}
    \includegraphics[width=7cm,height=8cm,angle=-90, bb= 50 115 554 680]{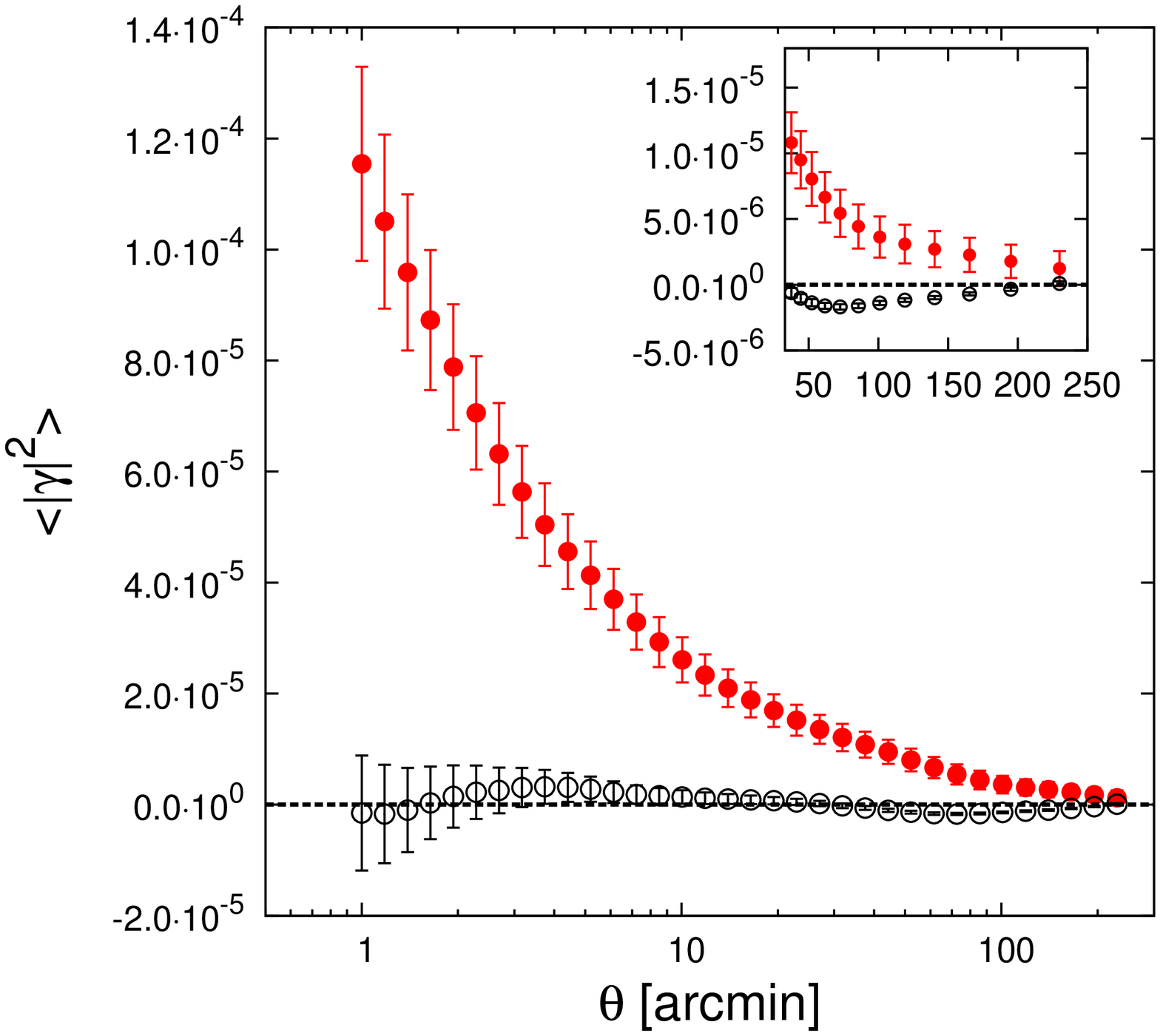}
    \caption{Two-point statistics from the combined 57 pointings.  The
       error bars of the E-mode include statistical noise added in
       quadrature to the non-Gaussian cosmic variance. Only
       statistical uncertainty contributes to the error budget for the
       B-mode. Red filled points show the E-mode, black open points
       the B-mode.  The enlargements in each panel show the signal in
       the angular range $35\arcmin$-$230\arcmin$.  }
    \label{shear}
  \end{center}
\end{figure}

\begin{figure}
\begin{center}
\hspace {-0.5cm}
\includegraphics[width=8.5cm]{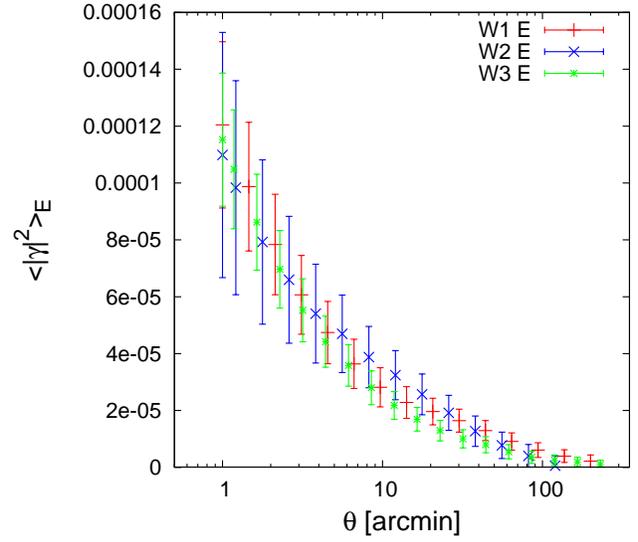}
\caption{The top-hat E-mode shear signals of W1 up to $200\arcmin$, of
 W2 up to $120\arcmin$ and of W3 up to $230\arcmin$ are shown. The
 error bars includes statistical noise and cosmic variance for each
 individual field.
}
\label{shearw123}
\end{center}
\end{figure}

The shear correlation functions $\xi_\pm$ are computed in narrow bins.
We use angular separations in the range between a conservative lower
limit of 3 arc seconds and a maximum separation $\vartheta_{\rm max}$
where the number of pairs per bin becomes very small. For each field,
the number of pairs per bin shows a similar ``top-hat" behavior: a
very steep increase from $\vartheta =0$ followed by a roughly constant
value up to a $\vartheta_{\rm max}$ where it starts a very steep
decrease to zero.  This separation $\vartheta_{\rm max}$ is
$400\arcmin$ for W1, $240\arcmin$ for W2, and $462\arcmin$ for W3.

From the two-point shear correlation functions we calculate the shear
top-hat variance and the aperture-mass dispersion up to a radius of
$\theta_{\rm max}$ which is half of the largest separation
$\vartheta_{\rm max}$, according to Eq.~(\ref{tophateb}) and
(\ref{mapeb}).

The missing information for $\xi_{\rm E, B}(\theta)$ and $\langle
|\gamma|^2\rangle _{\rm E, B}(\theta)$ on scales larger than
$2\theta_{\rm max}$ is accounted for by adding theoretical predictions
of these off-sets to the data using a fiducial cosmological model.
Alternatively, we may set the B-modes of the shear correlation
function and top-hat variance to zero on the angular scales where we
measure zero aperture-mass dispersion B-modes.  We checked that both
methods produce very similar and small off-set values and thus this
procedure does not bias the cosmological interpretation towards the
fiducial model used. Furthermore, our cosmological estimates are made
using the aperture-mass dispersion and are free of this small
arbitrariness.

 The three statistics are plotted in Fig.~\ref{shear} and the
corresponding values and errors are provided in
Table~\ref{2pttablexi}-\ref{2pttablemap} in Appendix B.  It is worth
noting that this is the first time that a cosmic shear signal has been
measured down to $i'_{AB}=24.5$, beyond scales of one degree . Notice
also that the independent measurements of the shear statistics made in
the three individual fields W1, W2, W3 are statistically consistent at
all scales. This is illustrated by Fig.~\ref{shearw123}, where the
three measurements of top-hat dispersion are shown.

In Fig.~\ref{shear} the error bars of the E-modes include
statistical noise and cosmic variance calibrated for non-Gaussianity,
while the error for the B-modes only includes statistical
uncertainty.  We find a clear E-mode signal and a B-mode which is
consistent with zero throughout the explored range of angular
scales, except between 50 and 130 arc minutes where there is a small
but significant feature in all three second-order functions. This bump
of the B-mode peaks at about 60-80 arc-minutes which are the side and
diagonal sizes of a Megacam field. We therefore guess it is due to a
correlation in PSF residuals on the scale of the camera.  In
Sect.~\ref{sec:cosmo} we show that our cosmological results are not
biased by this level of residual systematics on this range of angular
scales.

\begin{figure*}[!bt]
\begin{center}
\includegraphics[width=7.5cm,height=7.5cm]{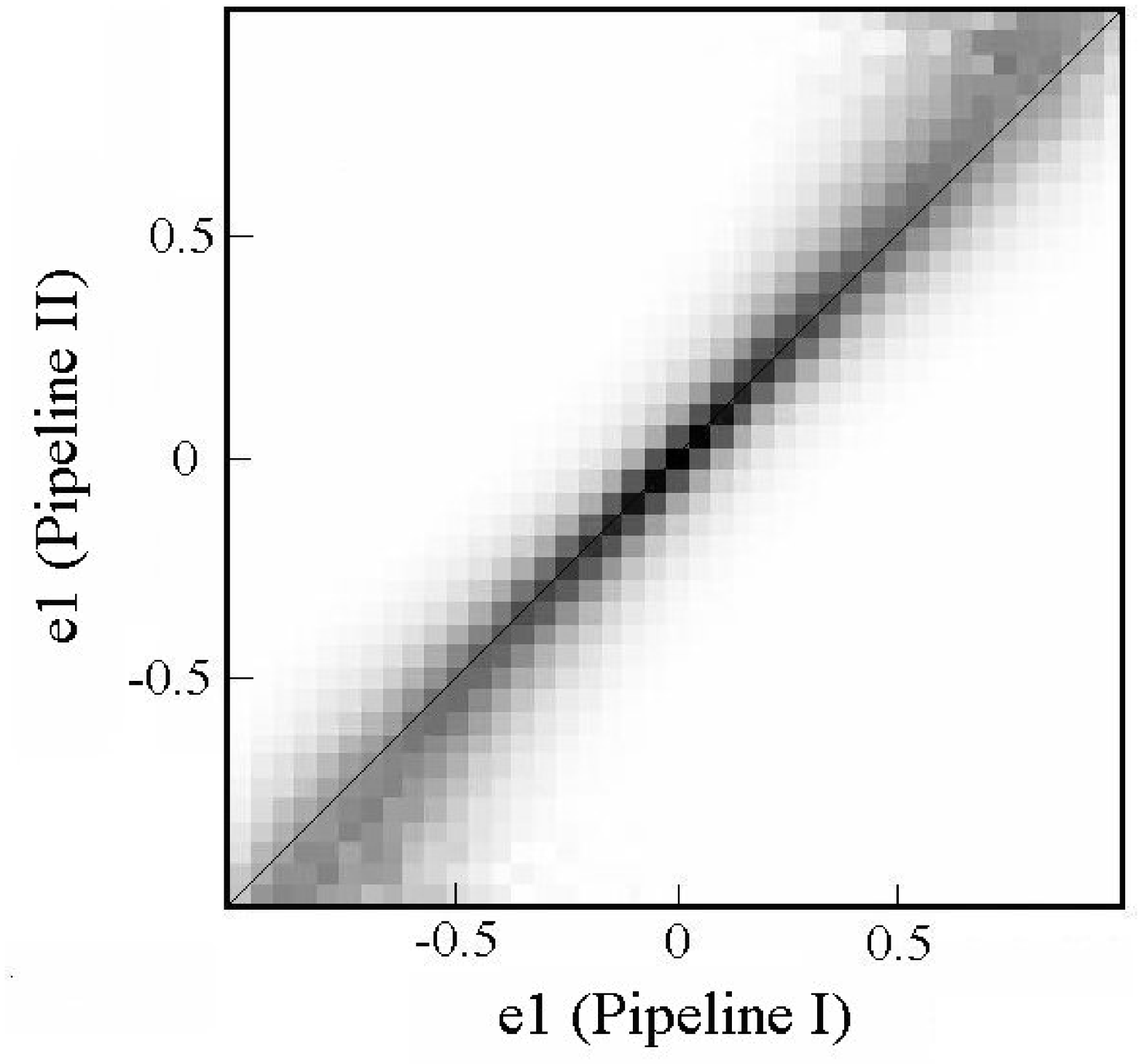}
\includegraphics[width=8.cm,height=7.5cm, bb = 80 50 370 300]{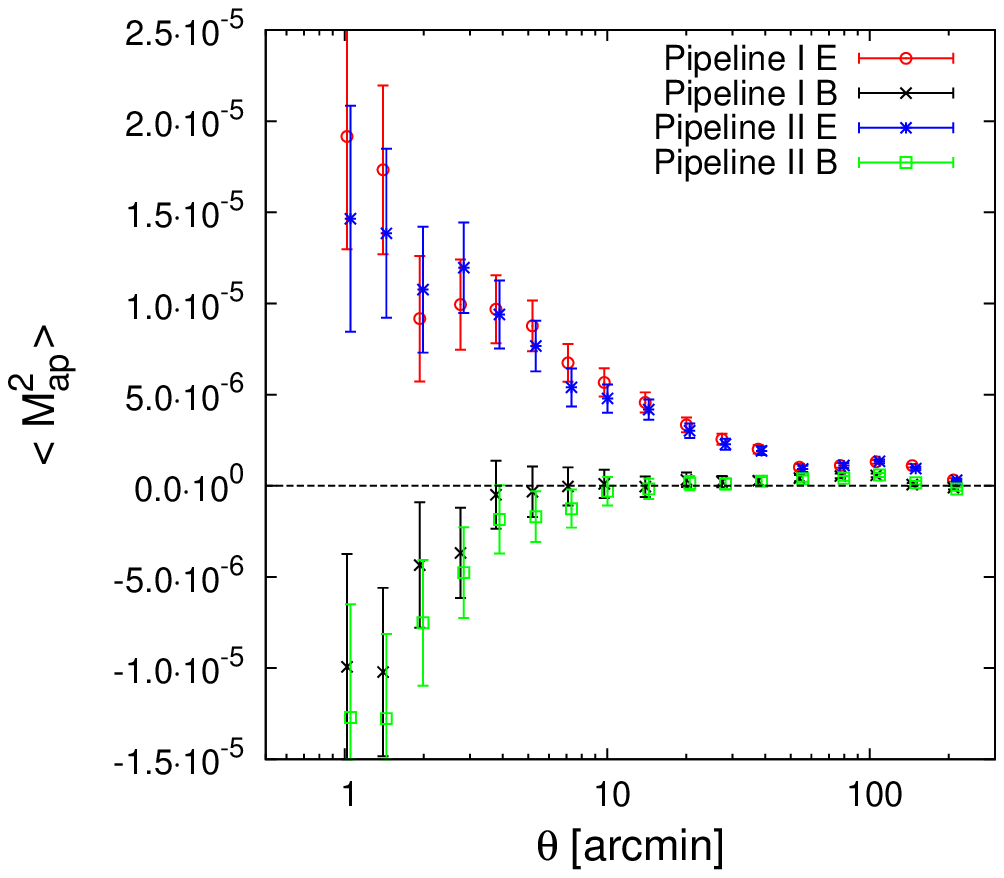}
\caption{\emph{Left panel:} Binned scatter plot of the shear estimates
  (one component) using the two pipelines. Dark colours show highest
  density of points. The bin size is 0.05 in $e_1$. \emph{Right
  panel:} The aperture-mass variance from W1, W2 and W3 measured with
  the two pipelines up to scale 210 arc minutes, using only objects
  which are detected in both pipelines. For clarity of the comparison,
  the error bars only show the statistical errors, but the
  cosmological analysis of this work includes the whole error budget
  (see text). These error bars are larger than the one of
  Fig.~\ref{shear} because number of common objects are smaller the
  full catalogue.  Note that the large negative B-mode on small scales
  is not present in the full catalogue, see Fig.~\ref{shear}.}
\label{com.map}
\end{center}
\end{figure*}

  On very large scales ($120'$-$230'$) we find a very small B-mode,
  much smaller than both the E-mode amplitude and cosmic variance, but
  which is not always within $1\sigma$ of a zero detection.  Notice
  that the errors on the B-mode shown in Fig.~\ref{shear} are
  theoretical (statistical) and not estimated from the data, which
  would include systematics (for example error contributions may arise
  from the incomplete PSF correction).  Moreover, the signal-to-noise
  with the present CFHTLS Wide data is so high, even for B-modes, that
  subtle effects may dominate the very small Poissonian error,
  particularly on large scales where there are a significant number of
  galaxy pairs.

   The field-to-field variation of the B-modes is a possible way to
   assess these effects on the error buget. We tried to measure this
   by splitting the 3 Wide fields into 11 blocks of 2$\times $2
   deg$^2$ each, which allows to calculate the B-modes on scales up to
   60 arcmin in each block. We obtained B-modes with amplitude very
   similar to Fig.~\ref{shear} but the field-to-field scatter is
   larger than the plotted error bars and reaches a factor of 2 at
   $60'$.  This is an interesting indication that we are likely
   underestimating the error on B-modes, even though it is not a
   precise measurement due to the small number of independant fields.
   A thorough analysis of this noise contribution needs many more
   field and is left to a future analysis of the CFHTLS four year
   data.

\subsection {Cross-check  and control of systematics}

We cross-checked the shear measurement by using an independent
analysis on the same data sets.  This analysis was done with another
version of KSB+ that has been tested with the STEP1+2 simulations
\citep[`HH' in][]{step1,step2}. Hereafter, we refer to our analysis as
`Pipeline I' and to the `HH' results as `Pipeline II'.

The left panel of Fig.~\ref{com.map} shows the shear estimated for
each galaxy by each of the pipelines.  The results are in good
agreement for ellipticity values per component between $-0.6$ and
0.6. For ellipticities outside this range the dispersion between the
pipelines is larger and a trend for an underestimation of the shear
from Pipeline I with respect to Pipeline II can be seen.  Note
however that the pipelines are not optimised for large ellipticities,
since the STEP simulation galaxies have ellipticities that are smaller
than 0.1.

We then compare the two-point functions using the aperture-mass
variance. We choose this statistic because angular scales are less
correlated than for the top-hat dispersion. Moreover, it does not have
any ambiguity related to a non-local E/B decomposition. The values of
$M_{\rm {ap}}$ are calculated from the two pipelines using only 
objects detected by both pipelines. Because the pipelines have
different selection criteria the common objects are only two-thirds of
the whole sample. Each object is assigned a weight which is the
product of its weights in each of the two pipelines.  The largest
radius explored in the comparison is 210 arc minutes.  As can be seen
in Fig.~\ref{com.map} (right panel), the E- and B-modes of the two
pipelines are remarkably similar. The differences are within the
$1\sigma$ errors on all angular scales.  The small B-mode bump
appears in both results at 60-80 arc minutes, as in
Fig.~\ref{shear}. It also drops to nearly zero at all scales beyond
120 arcmin for both pipelines.  The bias between the two pipelines at
large ellipticities, seen in the left panel of Fig.~\ref{com.map} is
not visible here. The reason is that the large ellipticity galaxies
represent less than 4\% of the sample.  Furthermore these galaxies are
typically downweighted; large ellipticities are difficult to measure,
resulting of a larger error on the ellipticity measurement, and the
shear polarisability increases with ellipticities. They have
consequently a lower weight according to Eq.~(\ref{weightelip}).

 These results are not expected to be identical to the aperture
mass dispersion computed with the whole sample, shown in
Fig.~\ref{shear}, because the number of objects in the two samples is
different. They are however similar, except for the large B-modes on
scales smaller than $2\arcmin$, which are detected by both pipelines
on the smaller sample. Since both analyses use KSB, these B-modes may
be due to similar residuals of the PSF correction, but we cannot rule
out an intrinsic B-mode contribution. Whatever the origin we only use
angular scales larger than $2\arcmin$ for the cosmological parameter
constraints (see also Sect.~\ref{sec:theo}) in order to avoid any
contamination.

The most common and problematic source of contamination of the lensing
signal is the imperfect PSF anisotropy correction. The angular
dependence of any PSF systematic residual may be checked by computing
the correlation between the corrected galaxy and uncorrected stellar
ellipticities.  Following \cite{2003MNRAS.344..673B} we normalise this
quantity by the star-star uncorrected ellipticity correlation in order
to assess its amplitude
\begin{equation}
\xi_{\rm sys}(\theta)=\frac{\langle e^{\star}(x)  \gamma(x+\theta)\rangle^2}{\langle e^\star(x) e^\star(x+\theta)
  \rangle},
\label{xisys}
\end{equation}
where the symbol $\star$ indicates a stellar quantity.  Figure~\ref{sys}
shows this cross-correlation compared to the shear signal $ \xi_{\rm
E}$ up to $230\arcmin$. Overall, the amplitude is at least one order
of magnitude smaller than the signal. This demonstrates that the PSF
correction is under control in our shear analysis up to an angular
scale of nearly 4 degrees.

\section{Source redshift distribution}

The calibration of the source redshift distribution in the the CFHTLS
Wide fields cannot be calculated from the Wide photometric data since
only a few fields have already been observed in 5 bands. However, the
CFHTLS Deep fields overlap, or are located very close to, the Wide
fields.  One can therefore use the photometric redshifts derived for
the CFHTLS Deep data \citep{Ilbert06} as a representative sample of
the Wide galaxy population, in particular for W1 that covers the D1
field.

\subsection{Building of the parent Deep $n(z)$ histogram}

The \cite{Ilbert06} catalogue samples photometric redshifts of more
than 500,000 objects in the four CFHTLS Deep fields\footnote {\tt
{http://terapix.iap.fr/rubrique.php?id\_rubrique=227}}, with an
$i'_{AB}$ limiting magnitude much fainter than that of the Wide
survey, covering the range $0 \leq z \leq 6$.  It has been calibrated
with spectroscopic redshifts obtained by the VVDS Survey in the CFHTLS
Deep D1 field \citep{Olf04}.  In this photometric redshift catalogue
318,776 galaxies have a magnitude matching the range used in the Wide
survey, i.e. $21.5 \leq i'_{AB} \leq 24.5$.  This sub-sample is used
to build up our redshift distribution.

For each object in \cite{Ilbert06}, the released photometric redshift
catalogue provides the maximum likelihood redshift $z_{\rm p}$ and
error estimates such as the left and right 1$\sigma$ error. In order
to estimate the redshift distribution we build a normalised Gaussian
probability distribution for each galaxy, with mean $z_{\rm p}$ and
dispersion given by the mean of the left and right error. We then draw
a redshift $z$ randomly and repeat the procedure 1000 times. The
variance of these 1000 randomizations is considered into the final
error budget.

\begin{figure}[!h]
\begin{center}
\includegraphics[width=8.cm,height=7.cm]{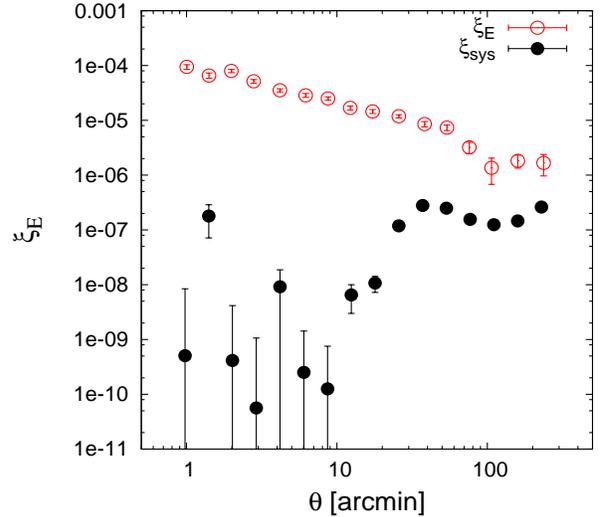}
\caption{The cross-correlation function $\xi_{\rm sys}$ (Eq.~\ref{xisys})
  between galaxies and stars is shown as a function of angular scale up
  to $230\arcmin$ (black filled). The amplitude of the cross-correlation is
  always at least one order of magnitude smaller than the shear signal
  $\xi_{\rm E}$ (red open).}
\label{sys}
\end{center}
\end{figure}
\subsection{Rescaling to the Wide population}

\begin{figure}[!t]
  \begin{center}
    \includegraphics[width=8cm]{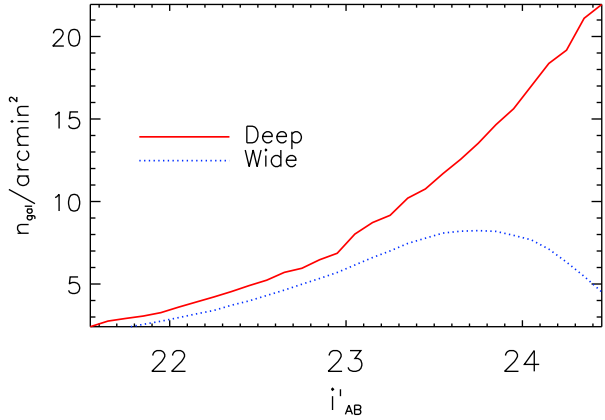}
    \caption{Magnitude distributions for the Deep (solid line) and
      Wide (dotted).  We use the ratio Wide/Deep for the rescaling of
      our redshift distribution. The Wide effective number density
      takes into account all weak lensing selection criteria and has
      been multiplied by their corresponding weights
      (Fig.~\ref{weight}).}
    \label{magdistribution}
  \end{center}
\end{figure}

To take into account the different selection functions between the
Deep parent sample and the Wide catalogue used in this work, each
galaxy is weighted according to the ratio of the Wide to Deep galaxy
number density, see Fig.~\ref{magdistribution}. In addition, we
include the weak lensing weight (Fig.~\ref{weight}) to match the
redshift distribution to the weighted galaxy population selected for
weak lensing. The redshift distribution is built up with all
photometric redshifts in the range $0 \le z \le 4$.

\subsection{Error budget}

The errors on the histogram have several contributions. First, the
uncertainty in the photometric redshifts is estimated from the
variance of the 1000 randomizations from the CFHTLS Wide redshift
histogram constructed in Sect.~5.1. Second, Poisson noise,
$\sigma_{\rm P}$ is added as $\sqrt{n}$, where $n$ is the number of
galaxies per redshift bin. Third, we need to include sample variance,
$\sigma_{\rm sv}$, since we estimate the redshift distribution from a
reference catalogue. The sample variance is given as a function of
Poisson noise and redshift for various survey areas in \cite
{vWW06}. We use the $\sigma_{\rm sv}/\sigma_{\rm P}$ ratio of a one
square degree survey, corrected for our bin size. We further rescale
it according to the weak-lensing selection function, since this
reduces the total number of galaxies, on which the ratio depends, as
$\sigma_{\rm sv}/\sigma_{\rm P} \propto n/\sqrt{n}$.  Note that we do
not divide the ratio by $\sqrt 4$ to account for having four
independent Deep fields, since the Poisson error is calculated for the
sum of the four fields.

With the large number of galaxies in our sample and the high accuracy
reached by the photometric redshifts at $z \sim 1$, the sample
variance is the dominating contribution to the error budget.  Poisson
noise and redshift uncertainties only contribute $\sim 5\%$ at $z=1$
but become dominant for $z>3$ where the number of galaxies is very
low. As a cross-check, we have calculated the field-to-field variance
of the four Deep photometric redshift catalogues. The result is
consistent with the sample variance obtained by \cite {vWW06}, using
numerical simulations.

\subsection{Fitting $n(z)$ of the Wide weak-lensing sample}
\label{sec:fitting_nz}

\begin{figure}[!t]
  \begin{center}
    \resizebox{\hsize}{!}{
      \includegraphics{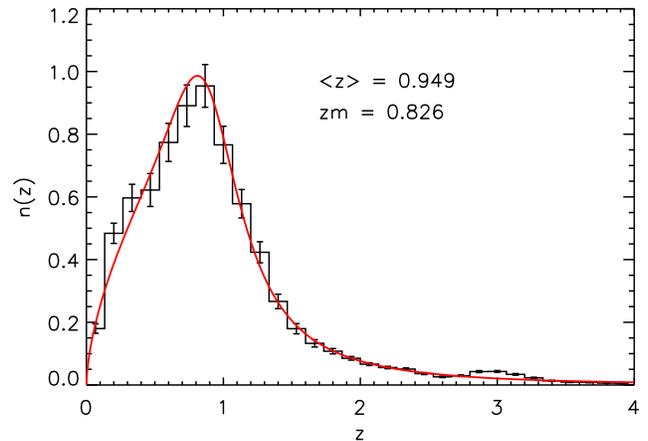}
    }
  \end{center}
  \caption{Final normalised redshift distribution. Galaxies are
    selected in the range [0;4], and the best-fit is given for
    function given in Eq.~(\ref{nz_power}). Note that the fit
    is only performed in the interval [0;2.5].  }
  \label{nz2ranges}
\end{figure}

  A histogram of the sources redshifts is shown in
 Fig.~\ref{nz2ranges}, where the error bars include redshift
 uncertainty, Poisson noise and sample variance. Although sample
 variance is taken into account, the histogram shows a significant
 bump at redshift $z\sim3$.  We cannot exclude the possibility that
 this small peak might be partly a real feature resulting from the
 joint spectroscopic, photometric and weak lensing selection functions
 of our galaxy sample.  It is however more likely to be an artifact
 due to systematic photometric redshift misidentifications arising
 from degeneracies that exist between the optical spectral energy
 distributions of galaxies with $z<0.2$ and $z>1.5$.  The recent
 analysis of the spatial correlation of populations in different
 photo-$z$ bins \citep[][ in prep.]{VW07} confirms that more than 50\%
 of galaxies in the peak are most probably at redshift $z \lsim 0.4$.

  We do not have a reliable estimate of the histogram bin-to-bin
  correlation.  Indeed, the off-diagonal sample variance was not
  calculated in the numerical simulation analysis of \cite {vWW06},
  and a field-to-field estimate using the four Deep fields is too
  noisy to be of practical and reliable use. Thus, in order not to
  propagate systematics present in the histogram into the cosmological
  constraints it is preferable to use a fitting function to the
  redshift distribution in the cosmological parameters estimation. For
  this we consider all galaxies in the range $0 \leq z \leq 
  2.5 $ and fit the redshift distribution with the following
  function,
\BE
n(z)= A\;\frac{z^a + z^{ab}}{z^b+c};\  A = \Big(\ \int_{0}^{z_{\rm
    max}}\frac{z^a + z^{ab}}{z^b+c}\,{\rm d}z\ \Big)^{-1} .
\label{nz_power}
\EE 
The normalization $A$ is determined by integrating until $z_{\rm
max}=6$, the upper limit of the photometric redshift catalogue.

 This function provides a better fit to the main peak and the tail of
the distribution as compared to the power-law function used in
B07. The distribution shown on Fig.~\ref{nz2ranges} corresponds to the
best-fit parameters listed in Table \ref{nzresults}. As expected, the
peak at $z\sim 3$ is no longer present. It is worth mentioning that,
although the histogram shows a significant fluctuation with respect to
the best-fit model at redshift $z\sim3$, the mean redshift derived
from the best-fit distribution is within 1\% of the mean value of the
histogram.

\begin{table}
  \caption{Results of the fit to the redshift distribution $0 \leq z
    \leq 2.5$, using Eq.~(\ref{nz_power}).  The $1\sigma$ error bars of
    three parameters are shown as well.
 $\langle z  \rangle$ is the mean, $z_{\rm m}$ the median redshift. }
  \label{nzresults}
  \begin{center}
    \newcommand{\rul}{\rule[-2.mm]{0mm}{5mm}}
    \begin{tabular}{cc|cc}
      \hline
      \hline
      \rul $a$ &  $0.612 \pm 0.043$  & $A$ &  $1.555$ \\
      \rul $b$ &  $8.125 \pm 0.871$  & $\langle z\rangle$ &  $0.949$ \\
      \rul $c$ &  $0.620 \pm 0.065$  & $z_{\rm m}$ & $0.826$ \\
      \hline
      \hline
    \end{tabular}
  \end{center}
\end{table}

\section{Cosmology with CFHTLS Wide}
\label{sec:cosmo}

\subsection{Shear covariance}
\label{sec:cov}

The covariance matrices for the shear two-point correlation functions
are calculated using the expressions of \cite{Schneider02}, valid for
a Gaussian shear field. They consist of a statistical noise term, a
cosmic variance term and a mixed term. To account for possible
residual systematics in the shear signal, we add the measured B-mode
at a given angular scale quadratically to the corresponding diagonal
element of the covariance.

The first three terms are calculated using a Monte Carlo method
applied to the measured galaxy positions and their weight similar to
the bootstrapping defined in Sect.~5. In that way the survey geometry,
boundary effects and the non-uniform, discrete galaxy distribution are
taken into account \citep{K04}.  Furthermore, this method allows to
compute a statistical noise that not only includes the shape noise of
the two-point functions estimators but also takes into account Poisson
or shot noise.

The non-Gaussianity of the shear field on small scales is considered
by applying a correction to the cosmic variance term using the
calibration formula of \cite{ESal07}.  The parameters for the model
shear correlation function are $\Omega_{\rm m} = 0.27, \Omega_\Lambda
= 0.73, h = 0.7, \Omega_{\rm b} = 0.044, \sigma_8 = 0.8$ and $n_{\rm
s} = 1.0$, using the \citet[][ hereafter S03]{Smith03} non-linear
prescription. The redshift distribution is the best-fit of the $n(z)$
data (see Sect.~5). For the non-Gaussian calibration a mean redshift
of 0.95 was assumed.

The top-hat variance, the aperture-mass statistic and the
E-/B-correlation functions are functions of both $\xi_+$ and $\xi_-$
(Eqs.~\ref{tophateb}-\ref{xieb}). Therefore, their covariance matrices
depend on the full covariance of the combined data vector ($\xi_+,
\xi_-$). However, we use only $C_{++}$, the covariance of $\xi_+$,
since the non-Gaussian calibration to the cosmic variance was derived
for this quantity \citep{ESal07}. We divide the Poisson term of
$C_{++}$ by two, which compensates for the additional information of
$\xi_-$. The other terms contributing to the total covariance
(mixed, Gaussian and non-Gaussian cosmic variance) do not depend on
the number of galaxy pairs per bin. Therefore, they are not affected
by not taking into account the Poisson-noise contribution from $\xi_-$
and thus they are unchanged.

\subsection{Parameter estimation}
\label{sec:parest}

\begin{figure*}[!th]
  \resizebox{\hsize}{!}{
   \includegraphics{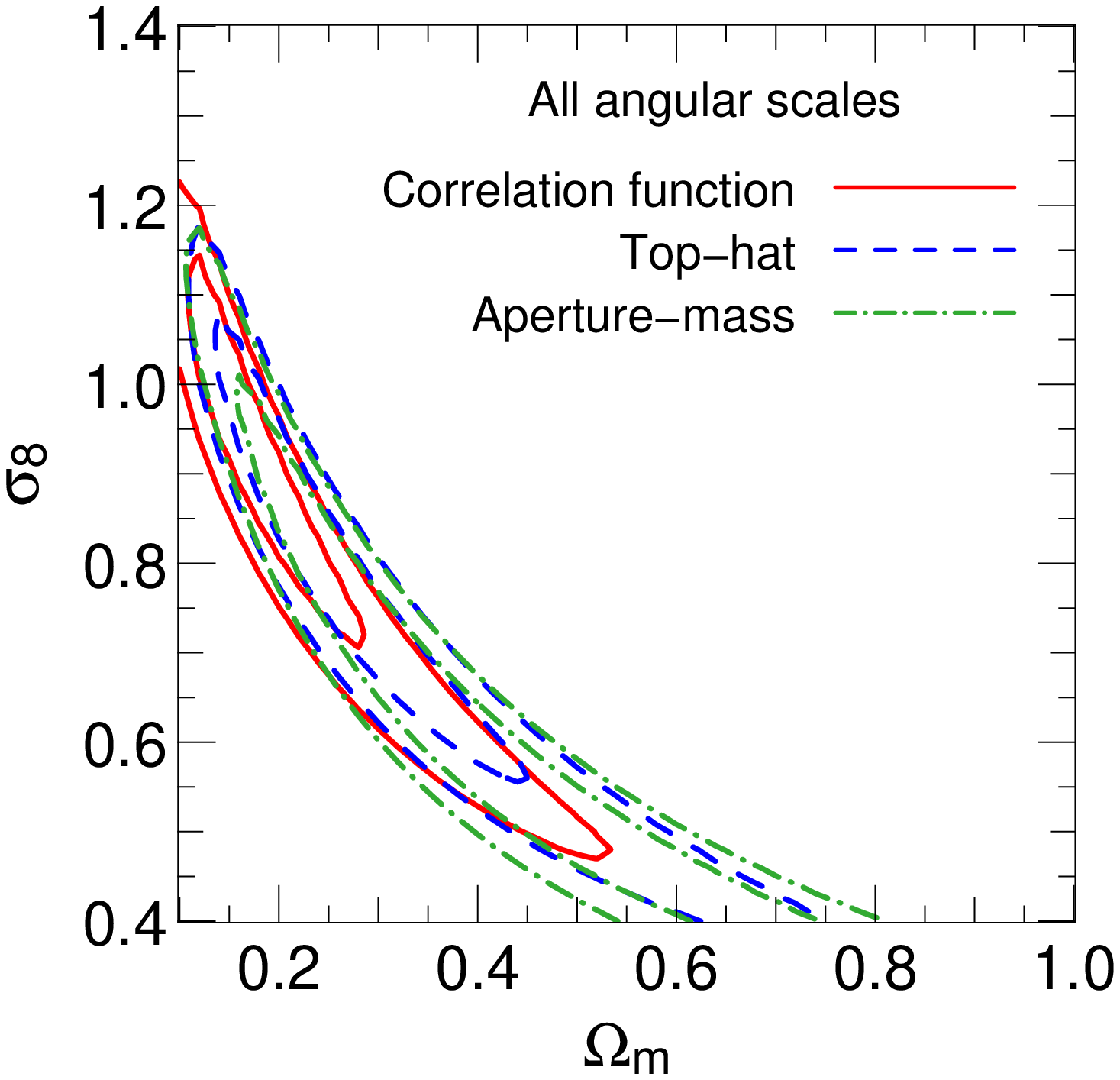}
    \includegraphics{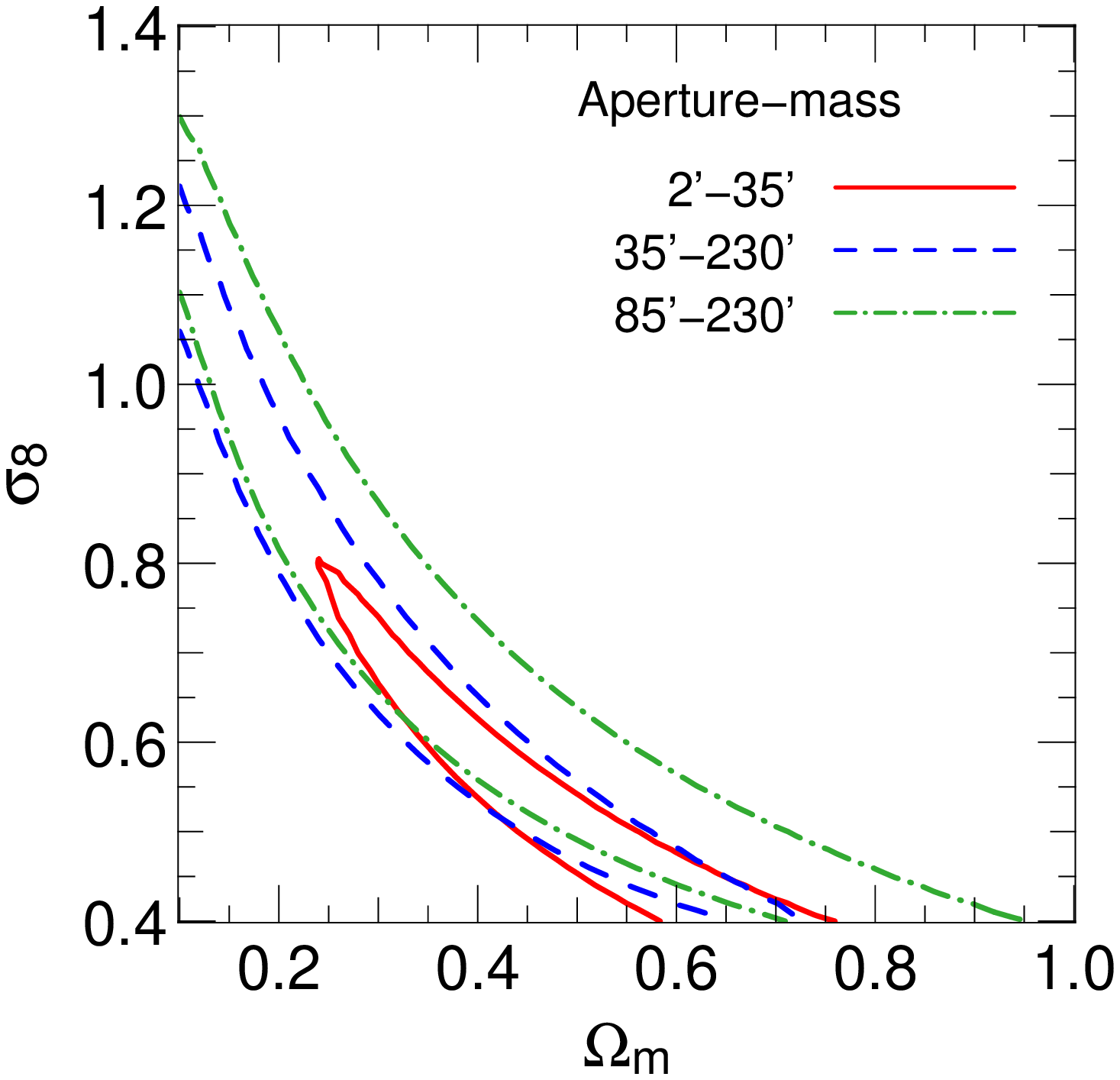}
  }

  \caption{\emph{Left panel:} Likelihood contours ($1,2\sigma$) for
    $\Omega_{\rm m}$ and $\sigma_8$, from the shear correlation
    function between 1 and 230 arc minutes (red solid lines), shear
    top-hat variance between 1 and 230 arc minutes (blue dashed), and
    aperture-mass dispersion between 2 and 230 arc minutes (green
    dotted-dashed). A flat, scale-free $\Lambda$CDM model with
    $\Omega_{\rm b}=0.044$ is assumed. We marginalise over $h$ and the
    redshift parameters.  \emph{Right panel:} 1$\sigma$ likelihood
    contours for $\Omega_{\rm m}$ and $\sigma_8$, from the
    aperture-mass variance between 2 and 35 arc minutes (red solid
    lines), for scales larger than 35 arc minutes (blue dashed) and
    for scales larger than 85 arc minutes (green dotted-dashed).  }

  \label{fig:Os-Sm-map-small+large}
\end{figure*}


The theoretical model that we fit to the data is a flat $\Lambda$CDM
cosmology with scale-free, adiabatic and Gaussian primordial
perturbations. The transfer function is the `shape fit' from
\cite{EH98} which takes into account baryonic suppression; we use a
fixed $\Omega_{\rm b} = 0.044$. The non-linear evolution of the power
spectrum is approximated with the fitting formula of S03.

The assumption of scale-invariance is not crucial for our results.
Indeed, marginalization over the primordial spectral index stretches
the confidence regions mainly along the $\Omega_{\rm m}$-$\sigma_8$
degeneracy direction. The obtained normalisation for a given
$\Omega_{\rm m}$ or the $\Omega_{\rm m}$-$\sigma_8$ relation remains
unchanged.

We calculate the log-likelihood on a grid of 6-dimensional parameter
space: three cosmological parameters $(\Omega_{\rm m}, \sigma_8, h)$
and three parameters of the redshift distribution $(a, b, c)$.

The Gaussian lensing log-likelihood is
\begin{equation}
  \Delta \chi^2 = \frac 1 2 \sum_{ij} \left( d_i - m_i \right) (C^{-1})_{ij}
  \left( d_j - m_j \right),
 \label{gridlike}
\end{equation}
where an element $d_i$ of the data vector is either one of the
measured $\xi_E(\theta_i)$, $\langle | \gamma|^2 \rangle_{\rm
E}(\theta_i)$ or $\langle M^2_{\rm ap} \rangle(\theta_i)$, and $C$ is
the covariance of the corresponding estimator. The model $m_i$ is the
theoretical prediction of the shear statistic for the same angular
separation $\theta_i$, and is a function of cosmological and redshift
parameters.

The grid intervals are $[0.1;1]$ for $\Omega_{\rm m}$, $[0.4;1.4]$ for
$\sigma_8$ and $[0.6;0.8]$ for the Hubble parameter $h$. The
redshift parameters values are taken inside of their $2\sigma$ range:
$[0.53; 0.69]$ for $a$, $[6.90; 10.2]$ for $b$ and $[0.49; 0.77]$ for
$c$.  Translated into extreme $\langle z \rangle$ values, this
corresponds to an exploration range of $[0.71;1.02]$.  Since the
three redshift parameters are correlated, the grid includes models
that should be rejected by the redshift likelihood alone. For this
reason we multiply the likelihood, Eq.~(\ref{gridlike}), by a prior
given by the likelihood of the redshift distribution estimation,
\begin{equation}
  \Delta \chi_{z}^2 = \frac 1 2 \sum_i \frac{\left( n_i -
      n(z_i)\right)^2}{\sigma_i^2}.
  \label{chi2nofz}
\end{equation}
Here $n_i$ is the (normalised) number of galaxies in the $i$-th
redshift bin of Fig.~\ref{nz2ranges} and $n(z_i)$ the fitting function
Eq.~(\ref{nz_power}), evaluated at the redshift bin centre.  The error
on $n_i$ is $\sigma_i$ the error bar of the histogram, we neglect the
cross-correlation between different bins.

\subsection{Constraints}
\label{sec:constr}

The left panel of Fig.~\ref{fig:Os-Sm-map-small+large} shows the
marginalised 2D-likelihood contours for $\Omega_{\rm m}$ and
$\sigma_8$ using the $n(z)$ of Table~\ref{nzresults}.  A fit to the degeneracy direction
yields
\begin{align}
 \sigma_8 (\Omega_{\rm m}/0.25)^{0.46} & = 0.784 \pm 0.049 &
 \text{for}
  \;\;\;\; \; &  \xi_{\rm E}; \nonumber \\
  \sigma_8(\Omega_{\rm m}/0.25)^{0.53} & = 0.795 \pm  0.042  & \text{for} \;\;\;\;\, &
  \text{$\langle |\gamma|^2 \rangle_{\rm E}$}; \nonumber \\
  \sigma_8 (\Omega_{\rm m}/0.25)^{0.64} & = 0.785 \pm  0.043 & \text{for} \;\;\;\;\, &
  \langle M_{\rm ap}^2 \rangle. \nonumber
\end{align}
The results for all three statistics are in excellent agreement.
Because of the E-/B-mode mixing \citep {K06} we do not use $\langle
M_{\rm ap}^2 \rangle$ on scales smaller than 2 arc minutes, therefore
we omit the first four data points for this statistic (see
Table~\ref{2pttablexi}-\ref{2pttablemap}). If we use the \citet[][
hereafter PD96]{PD96} non-linear power spectrum, the resulting
$\sigma_8$ is about 2\% larger than for S03 for a fixed $\Omega_{\rm
m} = 0.25$.

\subsection{Separating small and large scales}
\label{sec:constr_large}

\begin{figure}

  \resizebox{\hsize}{!}{
    \includegraphics{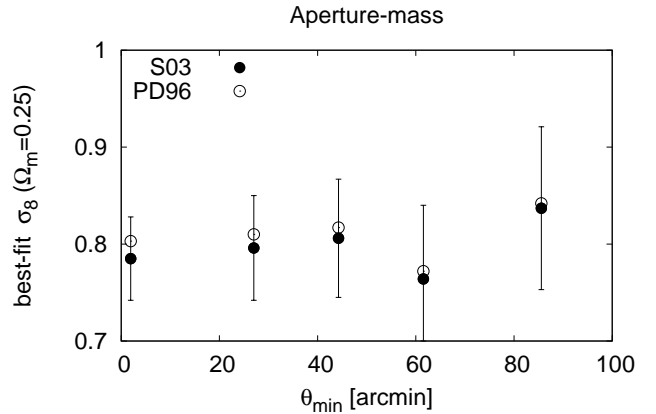}
  }

  \caption{The best-fit $\sigma_8$ as function of the minimum angular
    scale $\theta_{\rm min}$ that is used for the
    $\chi^2$-analysis. Results are shown for S03 (open symbols and
    $1\sigma$ error bars) and PD96 (filled symbols). The difference between
    the two non-linear models decreases at large scales.}
  \label{fig:nlbias}
\end{figure}

\begin{figure*}[!bt]

  \resizebox{\hsize}{!}{
     \includegraphics{8522f12a.ps}
     \includegraphics[scale=1.325]{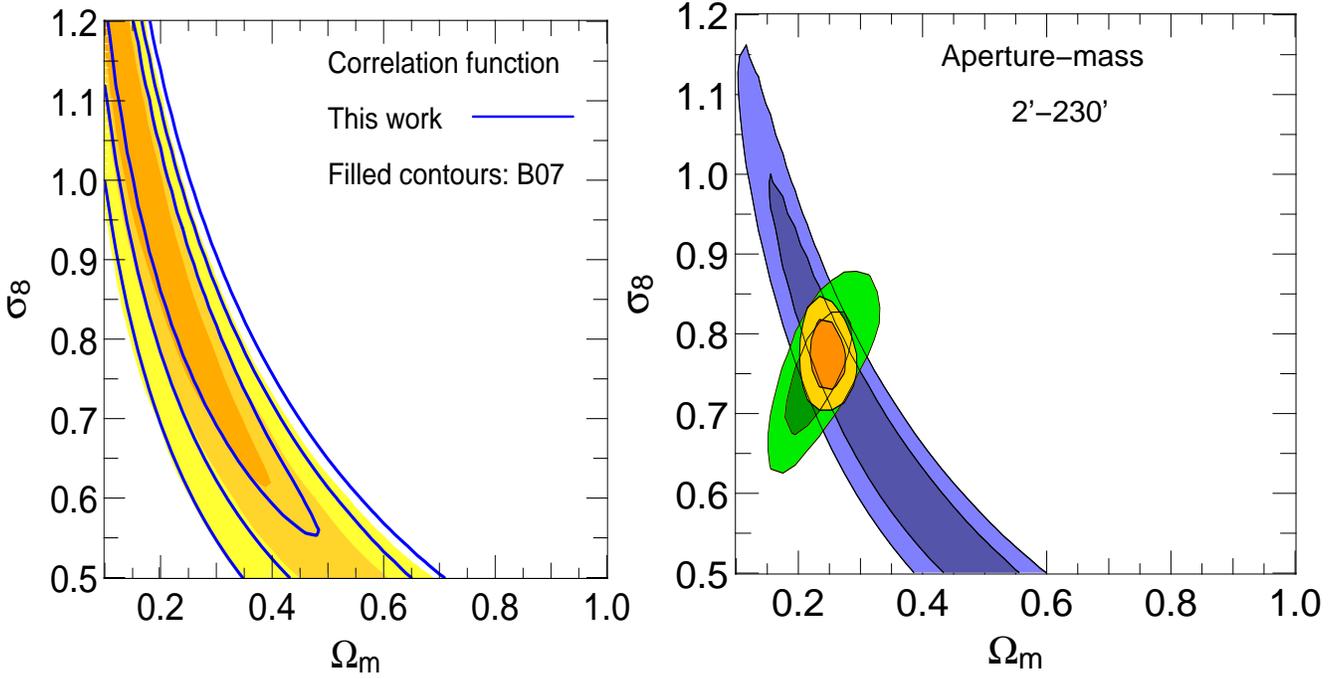}
  }

  \caption{\emph{Left panel:} Comparison ($1, 2, 3\sigma$) between our
    results (bold lines) and the 100 square degree survey (B07, filled
    contours), using $\xi_{\rm E}$ in both cases. The redshift
    distribution is fitted in the range of $[0.2;1.5]$ to be
    consistent with B07.  \emph{Right panel:} Comparison ($1,
    2\sigma$) between WMAP3 \citep[green contours,][]{Spergel07} and
    our $\langle M_{\rm ap}^2 \rangle$-results between 2 and 230 arc
    minutes (purple). The combined contours of WMAP3 and CFHTLS Wide
    are shown in orange.}
  \label{fig:JonBencomp}
\end{figure*}

Because of the large connected area of the CFHTLS Wide, we are able to
obtain interesting cosmological constraints by using large scales
only. Although the error bars increase when small scales are not taken
into account, the sensitivity to several systematic effects is
strongly reduced. The deviation from the linear prediction of the
shear top-hat dispersion is 20\% at a scale of $35\arcmin$, for the
redshift range probed by the Wide survey. The non-linear to linear
ratio of $\langle M_{\rm ap}^2 \rangle$ is 3 at $35\arcmin$ and 1.5 at
$85\arcmin$, respectively.  Our signal on large scales is therefore in
the linear regime and the resulting constraints do not depend on the
details of the non-linear modeling.  In particular, we are not
sensitive to the difference between PD96 and S03 as can be seen in
Fig.~\ref{fig:nlbias}.  Other systematics which might bias the results
on small scales are baryonic effects
\citep[e.g.][]{2004ApJ...616L..75Z}, intrinsic alignment and, maybe
most important, shear-shape correlations. All these effects are not
yet well understood as they depend on structure formation on small
scales and the relationship between galaxies and dark matter. In
particular, the shear-shape correlation leads to an underestimation of
$\sigma_8$ \citep{Hirata04,Hirata07}.  On scales larger than about
$10\arcmin$ the shear field is Gaussian. The non-Gaussian calibration
of the covariance matrix is not needed and also the Gaussian
assumption of the likelihood is justified. These two factors will
yield more accurate error estimates on the cosmological parameters.

In the right panel of Fig.~\ref{fig:Os-Sm-map-small+large} the results
for small and large scales are shown. By using only small scales we
obtain tighter constraints than by using only large scales, as the
signal-to-noise ratio is higher. Using the aperture-mass dispersion,
 the constraints derived from the three angular ranges are in
very good agreement, with all mean values within $1\sigma$:
\begin{align}
  \sigma_8 (\Omega_{\rm m}/0.25)^{0.66} = & \, 0.780 \pm 0.044 &  \text{for} \;\;\;\;\; &
  2'<\theta<35'; \nonumber \\
  \sigma_8 (\Omega_{\rm m}/0.25)^{0.54} = & \, 0.780 \pm 0.060 &
  \text{for} \;\;\;\;\;& 35'<\theta<230'; \nonumber \\
  \sigma_8 (\Omega_{\rm m}/0.25)^{0.53} = & \, 0.837 \pm 0.084 & \text{for} \;\;\;\;\;&
  85'<\theta<230' . \nonumber
\end{align}

These results are stable to changes in the smallest angular scale used.
For example, $\sigma_8$ changes by half a percent when only scales
larger than 4 arc minutes are used.

We checked that these constraints are not sensitive to possible
systematics on angular scales between 50 and 130 arc minutes, where
the B-mode shows a significant bump. We fit cosmological parameters
using scales with $2' < \theta < 50'$ plus $130' < \theta < 230'$, and
found the same results for $\Omega_{\rm m}$ and $\sigma_8$.  On the
other hand, fitting only the affected scales, $50' < \theta < 130'$,
we get $\sigma_8 = 0.840 \pm 0.063$ for $\Omega_{\rm m} = 0.25$, which
is consistent with the results from other scales.

\subsection{Comparison with other data sets}
\label{sec:comp}

\begin{figure}[!bt]
   \includegraphics[width=8.5cm]{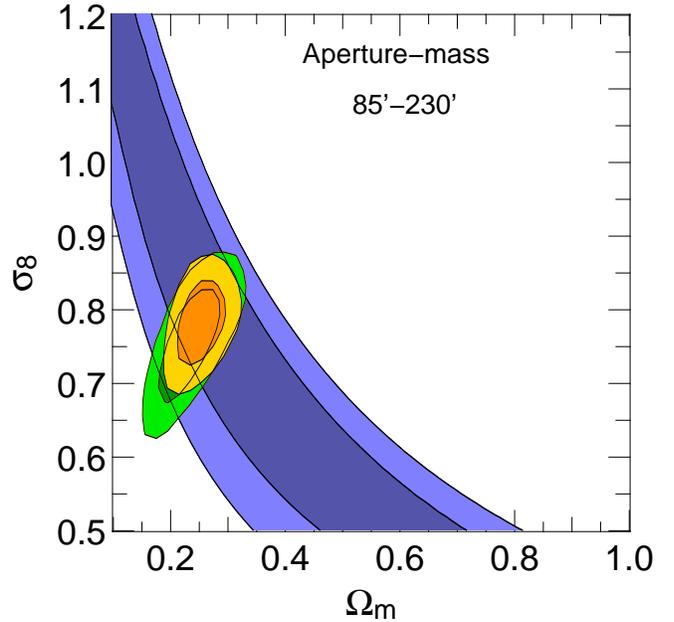}
  \caption{ Comparison ($1, 2\sigma$) between WMAP3 \citep[green
    contours,][]{Spergel07} and our $\langle M_{\rm ap}^2
    \rangle$-results in linear scale only ($85\arcmin$--$230\arcmin$,
    purple). The combined contours of WMAP3 and CFHTLS Wide are shown
    in orange.  }
  \label{fig:wmap3wllinear}
\end{figure}

\begin{table*}
\caption{The combined constraints from CFHTLS and WMAP3 for
$\Omega_{\rm m}$ and $\sigma_8$.}
 \label{tabsigma8}
 \begin{center}
 \begin{tabular}{cc||cc}
 \hline
 \hline
Two-point function & Angular scales &   $\Omega_{\rm m}$ & $\sigma_8$\\
\hline
$\xi_{\rm E}$&  $(1' < \theta < 230')$ & $0.243 \pm 0.020$ & $0.771\pm 0.030$ \\
$\langle |\gamma|^2 \rangle_{\rm E}$ & $(2' < \theta < 230')$ & $0.249 \pm 0.019$ & $0.776\pm 0.029$ \\
$\langle M_{\rm ap}^2 \rangle$  & $(2' < \theta < 230')$ & $0.248 \pm 0.019$ & $0.771\pm 0.029$\\
&&&\\
$\langle M_{\rm ap}^2 \rangle$ &  $(85' < \theta < 230')$ &  $0.255 \pm 0.027$ & $0.782 \pm 0.038$ \\
\hline \hline
\end{tabular}
 \end{center}
 \end{table*}

Our results on cosmological parameters are in very good agreement with
the most recent cosmic shear analysis which combined the first CFHTLS
Wide data release, the RCS, the {\sc VIRMOS-Descart} and the GaBoDS
surveys (the `100 square degree survey', B07).  In order to compare
the two results we construct a new Wide $n(z)$ histogram that has a
consistent redshift distribution. Following B07, we only use CFHTLS
Deep galaxies with a photometric redshift maximum peak probability in
the range $[0.2;1.5]$. We fit an exponential function proposed by
\cite{1993MNRAS.265..145B} and \citet{vWM02} in the same $z$-range.
The mean redshift $\langle z \rangle = 0.792$ matches the one in B07.
The left panel of Fig.~\ref{fig:JonBencomp} shows an excellent
agreement between the two results. The comparatively smaller sky
coverage of our survey is compensated by its larger range of angular
scales.  It is also interesting to notice that our results are in
excellent agreement with the CTIO survey
\citep{2003AJ....125.1014J,JBBD06}.

Next, we compare our results for $\Omega_{\rm m}$ and $\sigma_8$ with
the Wilkinson Microwave Anisotropy Probe 3-year constraints
\citep[WMAP3,][]{Spergel07}. We combine our likelihood with a CMB one
computed for a flat $\Lambda$CDM cosmology using WMAP3 data only
including temperature (TT), temperature-polarisation (TE) and
polarisation (EE) modes. The combination of the two data sets leads to
remarkably smaller confidence levels as compared to individual
ones. In particular, as shown in the right panel of
Fig.~\ref{fig:JonBencomp}, the combination of CFHTLS using the
aperture-mass variance and WMAP3 breaks the severe $\Omega_{\rm
m}$-$\sigma_8$ degeneracy. This translates into a reduction of the
region allowed with 95$\%$ confidence level by a factor of 3.15 as
compared to WMAP3 only.  The marginalised constraints for each
parameter are shown in Table~\ref{tabsigma8}.  This corresponds to a
relative accuracy of $8\%$ in $\Omega_{\rm m}$ and $4\%$ in
$\sigma_8$, improving the WMAP3 constraints of \citet[][
Table~5]{Spergel07} by a factor of 1.82 and 1.77 respectively. The
combinations of CFHTLS and WMAP3 using the shear correlation function
and top-hat shear variance show consistent results for $\Omega_{\rm
m}$ and $\sigma_8$ as listed in Table~\ref{tabsigma8}.

In view of the weak lensing signal we found on large scales, we
combine the WMAP3 data with the CFHTLS beyond one degree only, and
examine the cosmological constraints derived from the linear
regime. We look at the constraints on $\Omega_{\rm m}$ and $\sigma_8$
by separating the large angular scales ($85\arcmin$--$230\arcmin$)
from the whole sample, which is listed in Table~\ref{tabsigma8}.  They
are shown in Fig.~\ref{fig:wmap3wllinear}.  One can see that the large
angular scales alone have a significant contribution to the total
constraint, although the survey only covers 57 deg$^2$.  It is then
realistic to predict from this figure that weak lensing surveys may
soon be able to explore cosmological models using linear theory only,
similar to CMB physics of primary anisotropies.  This is very
promising for future surveys with sky coverage  much larger
than CFHTLS Wide at the same depth.  Equivalent constraints from the
linear structures, similar to the ones shown in
Fig.~\ref{fig:wmap3wllinear} will then be narrower by a factor of at
least 10.

Our joint analysis with WMAP3 data is in full agreement with
similar studies presented in \cite{Spergel07}, using several other
data sets.  Our estimate for the matter density also coincides with
the result derived by \cite{Astier06} based on their SNIa light curves
only, for a flat $\Lambda$CDM Universe. The comparison with clusters
of galaxies is, in contrast, less conclusive.  Cluster observations
estimate a broad range of $\sigma_8$ values, with some being fully
consistent with our results
\cite{Gladders07}, (see also \cite{Hetterscheidt07} for a compilation
of results), while a recent analysis of simulations argue for higher values
\citep{Evrard07,Yepes07}.  The trends for a high value of  $\sigma_8$ are
also derived from analyses of the Lyman-alpha forest \citep[see][ and
reference therein]{Slosar07}.

\section{Contamination by  shear-shape correlation}

   The gravitational lensing signal may be contaminated by the
   intrinsic alignment and by the gravitational shear and intrinsic
   ellipticity (or shear-shape) correlations.  We do not consider the
   first term since it would be negligible due to a broad redshift
   distribution of our sample. On the other hand,
   \cite{2006MNRAS.367..611M} and \citet{Hirata07} pointed out that
   the shear-shape anti-correlation may bias the estimate of
   $\sigma_8$ by 1 to $20 \% $ for a $ \langle z \rangle =1$ survey on
   angular scales that we have explored in this work. It is therefore
   important to estimate its amplitude and to which extent it may
   spoil our cosmological constraints. We

We attempt a rather simple analysis of the shear-shape correlation
 (GI) contribution to the shear signal. We use the following simple
 model for the GI correlation function $\xi_{\rm GI}$, which is
 motivated by numerical simulations \citep{Heymans06}
\begin{equation}
  \xi_{\rm GI}(\theta) = {\cal E} \frac{A}{\theta + \theta_0}.
\end{equation}
The lensing efficiency $\cal E$ is weighted by the source redshift
distribution 
\begin{equation*}
  {\cal E} = \int\limits_0^{\chi_{\rm lim}}  {\rm d} \chi_{\rm l} \, n(\chi_{\rm l})
   \int\limits_{\chi_{\rm l}}^{\chi_{\rm lim}} {\rm d} \chi_{\rm s} \,
   n(\chi_{\rm s})
  \frac{f_K(\chi_{\rm l}) f_K(\chi_{\rm s}-\chi_{\rm l})}{f_K(\chi_{\rm s})}.
\end{equation*}
For our fiducial flat model with $\Omega_{\rm m} = 0.25$
 and the redshift distribution of Table~\ref{nzresults}, we obtain
${\cal E} = 95.54 \, \text{Mpc}/h$. We fix the scale $\theta_0$ to 1 arc
minute, and further set $\xi_{\rm GI} = 0$ on scales larger than 1
degree.

\begin{figure}[!tb]

  \begin{center}

    \hspace*{-4em}%
    \includegraphics[height=8cm]{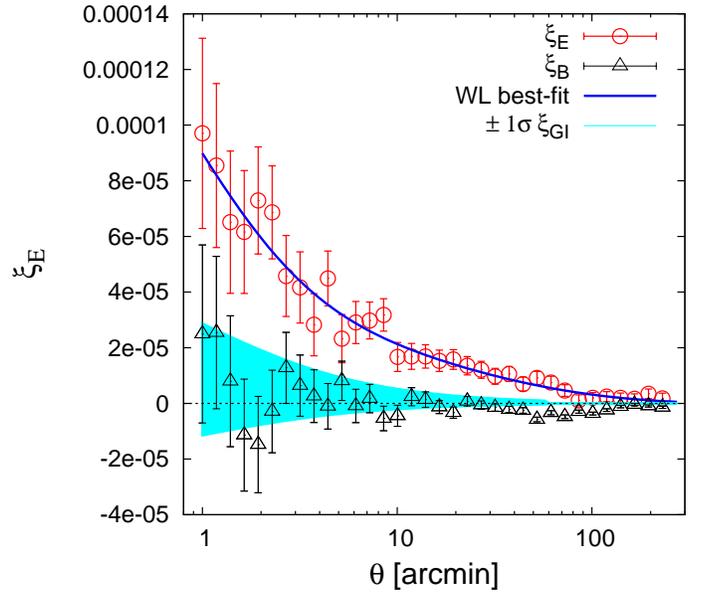}
  \end{center}

  \caption{ The measured $\xi_{\rm E}$ and $\xi_{\rm B}$ (open symbols
and error bars) with the lensing-only best-fit curve (solid blue line)
and the allowed fractional $\pm 1\sigma$-contribution of $\xi_{\rm
GI}$ to the total signal (shaded cyan region).}
  \label{fig:GI}
\end{figure}

We perform a combined likelihood analysis using the six cosmological
parameters as described in Sect.~\ref{sec:parest} and the GI amplitude
$A$. The sum $\xi_{\rm E} + \xi_{\rm GI}$ is fitted to the data. Since
the 7D-likelihood analysis is very time-consuming, we use the
marginalised 2$\sigma$ likelihood-region from the pure lensing
analysis (Sect.~\ref{sec:constr}) as a flat prior and do not consider
models outside this region. The marginalised result on $A$ is
consistent with zero.  We find for the amplitude $A$ in units of
$[10^{-7} h/$Mpc arcmin],
\begin{align}
  A & =  2.2^{+3.8}_{-4.6} & \text{for} \;\;\;\;\; & 1\arcmin <
  \theta < 230\arcmin , \nonumber
\end{align}
where the error indicates the $68\%$ confidence region.
 Figure~\ref{fig:GI} shows there is no significant signal detected at
 any scales.  The positive (negative) limit from all scales imply a
 +32\% (-13\%) contamination of the total signal by GI at one arc
 minute.

 Although the confidence region for the constrained GI amplitude is
 large it favours positive correlations, whereas from theory we would
 expect the GI signal to be negative \citep{Hirata04}.  As a
 consistency check we used a cosmology prior given by the marginalised
 $1\sigma$ likelihood region from a pure lensing analysis of the large
 scale results with $\theta > 60$ arc minutes.  The model $\xi_{\rm E}
 + \xi_{\rm GI}$ is then fitted on scales with $\theta < 60$ arc
 minutes. The resulting marginalised likelihood for $A$ favours
 negative GI models but is still consistent with zero.  This ansatz
 gives a high weight to the large-scale cosmic shear signal, and any
 systematics still present will influence the result.  The large scale
 increase in the measured star-galaxy cross correlation shown in
 Fig.~\ref{sys} highlights this concern.  As we cannot currently
 distinguish between GI and other possible systematic effects we can
 only conclude from our simple analysis that we find no evidence for a
 non-zero GI signal.

If our galaxy sample is strongly dominated by high-redshift spiral
galaxies, then the GI signal may be considerably weakened, as one can
anticipate from the morphological analysis of \cite
{2006MNRAS.367..611M}. We do not have enough colour data to explore in
detail the spectral/morphological types of the galaxies used in this
work. However, \cite {Zucca06} pointed out that about 80\% of the VVDS
spectroscopic galaxy sample up to $i^{\prime}_{AB}=24$ is composed of
spiral-like galaxies. It is then possible that the fraction of spirals
is much higher than elliptical galaxies in the population we are
sampling with cosmic shear. If so, it would reduce the contamination
to a very small effective contribution \citep{Heymans06}. A more
detailed investigation of the shear-shape analysis using photometric
redshifts and spectrophotometric information of galaxies is therefore
needed and will be discussed in a forthcoming paper.

\section{Summary and conclusions}

We have presented the weak lensing analysis of the CFHTLS T0003 Wide
data. The survey covers 57 deg$^2$, about two times the size of the
previous analysis by \citet{HH06}, and includes a new independent
field W2.

The galaxy shape measurements of a homogeneous sample of two million
galaxies down to $i'_{AB}=24.5$ have been validated using the STEP1
and STEP2 simulations \citep{step1,step2}. The top-hat shear variance,
aperture-mass dispersion and the two-point shear correlation functions
show a significant signal, with no galaxy-star correlations, from 1
arc minute up to 4 degrees.  The B-mode is consistent with zero on
most of these angular scales. It shows, however, a statistically
significant feature in the range 50-130 arc minutes, of unknown
origin. We have verified that this feature does not influence the
cosmological results.

 The two-point statistics show all expected properties of a cosmic
shear signal up to angular scales 10 times larger than the largest
non-linear scales of the survey.  Hence, for the first time the cosmic
shear signal can be explored with enough confidence to physical scales
of about 85 Mpc assuming lenses at $z=0.5$, for a flat Universe with
$h=0.72$ and $\Omega_{\rm m}=0.27$.  This is by far the widest scale
ever probed by weak lensing at that depth.

The weak lensing Wide data and the photometric redshifts sample of
\citet{Ilbert06} are both part of the CFHTLS T0003 release and cover
common fields. The redshift distribution of the Wide data can
therefore be calibrated using these photometric redshifts, assuming
with a high confidence level that the two galaxy populations are
similar.  Taking into account the selection criteria of the weak
lensing sample, we find a mean redshift of $\langle z\rangle = 0.949$
and a description of the redshift distribution in excellent agreement
with B07.

Using this redshift distribution, an exploration of constraints on
$\Omega_{\rm m}$-$\sigma_8$ has been carried out within the angular
range $1'\leq\theta\leq230'$.  The marginalised result on $\Omega_{\rm
m}$-$\sigma_8$ derived from the aperture-mass variance
\begin{align}
  \sigma_8 (\Omega_{\rm m}/0.25)^{0.64} & = 0.785 \pm  0.043 \ , &  \;\;\;\; &
   \nonumber
\end{align}
  is in excellent agreement with those obtained by the two other
  statistics (see Fig.~\ref{fig:Os-Sm-map-small+large}).  These
  constraints perfectly match those from B07 as shown in the left
  panel of Fig.~\ref{fig:JonBencomp}.  This is interesting because the
  two samples result from complementary approaches: the B07 sample
  explores the consistency of weak lensing results obtained from a
  heterogeneous sample consisting of four surveys. In contrast, our
  work analyses a very homogeneous data set consisting of one single
  survey, and using photometric redshifts derived from the same CFHTLS
  release within the same fields as the CFHTLS Wide.

 There is a clear trend towards a lower $\sigma_8$ as compared to
\citet{ESal06} and \cite{HH06}. This is a result of the less accurate
redshift distributions used in these analyses which were estimated
from the Hubble Deep Field photometric redshift sample.  This is well
confirmed when we combine our predictions on $\Omega_{\rm m}$ and
$\sigma_8$ with WMAP3 of \citet{Spergel07}, shown in the right panel
of Fig.~\ref{fig:JonBencomp}. There is a striking difference with
respect to the early comparison done by \citet{Spergel07}, using the
CFHTLS T0001 results. The $1.5\sigma$ tension is no longer visible; in
contrast, there is a large overlap between the two data sets.  The
joint CFHTLS-WMAP3 likelihood analysis then leads to tight
marginalised constraints on $\Omega_{\rm m}$ and $\sigma_8$,
\begin{equation*}
  \Omega_{\rm m}=0.248 \pm 0.019 \ \ {\rm and} \ \ \sigma_8=0.771\pm 0.029 \ ,
\end{equation*}
corresponding to an accuracy of  8\% and 4\% on these two
parameters. Hence, using a much better photometric redshift sample,
 based on the Deep CFHTLS T0003  data sets
 that directly
calibrate the genuine CFHTLS galaxy population, removes one of
the primary uncertainties of earlier CFHTLS  weak lensing analysis.

Considering the potential nuisances of systematic effects related to
non-linear scales, we split the sample into three ranges of angular
scales: the `highly non-linear' ($2'\leq \theta \leq 35'$), the
`intermediate' ($35'\leq \theta \leq 230'$) and the `linear'
($85\arcmin \leq \theta \leq 230\arcmin$) scales.  The analysis of the
three sub-samples do not reveal significant differences between each
regime (see Fig.~\ref{fig:Os-Sm-map-small+large}, right panel).  The
results are also stable to changes in the lower angular scales
increasing from $2'$ to $4'$.  This shows that the CFHTLS Wide cosmic
shear survey is not yet dominated by uncertainties related to our poor
knowledge of astrophysical systematics at small scales.  Finally, we
find that excluding scales with a significant B-mode ($50'\leq \theta
\leq 130'$) from the analysis does not change our results.  The
constraints on $\Omega_m-\sigma_8$ are therefore insensitive to the
level of residual systematics in our data.  All these tests strengthen
the confidence and reliability of our results.

The very large range of angular scales explored by the CFHTLS Wide
 opens a new window to cosmic shear surveys.
   It enables for the first time a comparison of cosmic shear and
 WMAP3 signals using only linear scales.  The constraints  shown in
Fig.~\ref{fig:wmap3wllinear} demonstrate
that there is still great predictive power from the linear regime
only. Future weak lensing surveys which cover areas significantly
larger than the CFHTLS will be able to pin down a much narrower region
in parameter space. Thus, it will be possible to obtain cosmological
parameters to percent-level accuracy and below from combining CMB and
weak lensing using  \emph{linear theory}.

Finally, the impact of the contamination by the shear-shape
correlation on cosmic shear surveys like CFHTLS is still unclear.  We
find its amplitude to be very low and compatible with zero at all
scales we explored. 
The low amplitude derived from
\cite{Hirata07}, using a survey shallower than the CFHTLS Wide, 
had already suggested that it should be a small effect and a
 difficult-to-detect signal in the CFHTLS-Wide, 
in particular if our galaxy sample is dominated by high-redshift
spiral galaxies \citep {Zucca06}.  At present, we can measure
$\sigma_8$ to a precision of about $5 \%$ and so  this bias is still
reasonably low.   With future work, however, this bias may
become the main source of error.

The CFHTLS is still in progress and the next release will include more
sky coverage and also a new field, W4. In this work, we only use the
wide $i'$-band data together with the photometric redshift from the
Deep T0003 $u^*,g',r',i',z'$.  The next step is therefore a more
detailed analysis of multi-colour data sets.  A better check of
systematics will be possible by cross-correlating the lensing signal
obtained independently in indifferent filters.  The larger CFHTLS Wide
sample with $u^*,g',r',i',z'$ will also improve tomographic studies
and will provide a photometric redshift to each individual galaxy.  We
will then be in the position to better control contaminations by
intrinsic alignment and the shear-shape (GI) correlations
\citep{Bridle06} and to move towards a full tomographic exploration of
the CFHTLS Deep and Wide surveys together.

{\acknowledgements We warmly thank  the
  CFHT, Terapix and CADC staff for their assistance and the
  considerable work they do to produce the CFHTLS data,  and the
  VVDS consortium for providing the galaxy spectroscopic sample in the
   CFHTLS fields. This work has
  made use of the VizieR catalogue access tool, CDS, Strasbourg,
  France.  We thank R.~Massey and the STEP collaboration for producing
  the STEP simulations used in this analysis, and Caltech, the
  University of British Columbia and JPL for their support to STEP. We
  thank F.~Bernardeau, T.~Erben, B.~Fort, M.~Hetterscheidt,
  H.~Hildebrandt, C.~Schimd, P.~Schneider, T.~Schrabback-Krahe,
  U.~Seljak, C.~Shu, J.-P.~Uzan for useful discussions.  LF thanks the
  ``European Association for Research in Astronomy" training site
  (EARA) and the European Commission Programme for the Marie Curie
  Doctoral Fellowship MEST-CT-2004-504604. ES aknowledges the
    support from the Alexander von Humboldt Foundation.  JC, LF, MK
  and YM thank the CNRS-Institut National des Sciences de l'Univers
  (INSU) and the French Programme National de Cosmologie (PNC) for
  their support to the CFHTLS cosmic shear program.  IT and YM
  acknowledge the support of the European Commission Programme
  6$^{th}$ framework, Marie Curie Training and Research Network
  ``DUEL'', contract number MRTN-CT-2006-036133. IT thanks the
  Deutsche Forschungsgemeinschaft under the project SCHN 342/8--1 and
  the Priority Programme 1177.  MK is supported by the CNRS ANR
  ``ECOSSTAT'', contract number ANR-05-BLAN-0283-04.  LVW, HH and MH
  are supported by the Natural Sciences and Engineering Research
  Council (NSERC), the Canadian Institute for Advanced Research (CIAR)
  and the Canadian Foundation for Innovation (CFI).  CH acknowledges
  the support of the European Commission Programme 6$^{th}$ framework,
  Marie Curie Outgoing International Fellowship, contract number
  M01F-CT-2006021891.  }


\begin{appendix}

\newpage

\section{STEP simulation calibration}

One of the crucial issues for weak lensing studies is the reliability
of galaxy shape measurement and the control of systematics.  The
detection and measurement of weak lensing is a technical
challenge. Weak distortion induced by gravitational lensing in the
observed shapes of galaxy images is only $\sim 1\%$, much smaller than
the typical intrinsic ellipticity dispersion $\sim 30\%$. To further
complicate the situation the observed shape of the galaxies is
affected by the PSF.  The Shear TEsting Programme\footnote{\tt
http://www.physics.ubc.ca/heymans/step.html} \citep {step1,step2},
hereafter STEP, is a collaborative project aiming to calibrate and
improve weak lensing methods using realistic Wide field simulated
images.  The first and second generation of STEP simulations
(hereafter STEP1 and STEP2) are designed for a ground-based survey.
In order to check the reliability of the shear measurement used in
this analysis, we calibrated the pipeline using all data sets from
STEP1 and STEP2.

STEP1 simulations contain relatively simple galaxy morphologies
generated using the SkyMaker software\footnote{\tt
http://terapix.iap.fr/soft/skymaker}. Five constant shears,
$\gamma_1^{\rm{input}} = $[0.0, 0.005, 0.01, 0.05, 0.1], are applied
to the galaxies, while the second component $\gamma_2^{\rm{input}}$ is
always set to zero.  Finally, galaxy and stellar point sources are
convolved with six different constant PSFs which attempt to reproduce
PSF shapes, that are typical of ground-based observations.  In this
way 30 sets of images, differing in PSF type and/or shear strength are
produced. Each set is composed of 64 images. The sky noise is
spatially uncorrelated.

STEP2 simulations contain complex galaxy morphologies produced using a
shapelet simulation package \citep {Massey04}.  Six sets of 64 images
with random constant input shears are convolved each with a different
optical PSF.  The six PSFs are chosen to span a range of realistic
ground-based observing conditions.  For each image, a twin image is
produced, in which galaxies are rotated by $90^o$ before applying the
same shear and the same PSF.  Combining the shear analysis on rotated
and non-rotated images demonstrates the pure measurement bias, since
the noise due to the scatter in a galaxies' intrinsic morphology is
removed.  The model of the sky noise is also more complex than the one
adopted to generate STEP1 simulations.  It is in fact a correlated
noise which aims to reproduce the noise of the drizzling process.

Our pipeline is an application of the KSB+ method. The observed galaxy
shape is modeled as a convolution of the sheared galaxy with the PSF,
which in turn is modeled as a circular profile convolved with a small
anisotropy.  Assuming the mean of the intrinsic ellipticity
distribution of galaxies to be zero and the PSF anisotropy to be
small, the first-order of the shear, $\gamma$, can be computed from
the observed ellipticities of galaxies, $\bd{e}^{\rm obs}$ as follows:
\begin{equation}
\bd{\gamma}=\langle \ P_{\gamma}^{-1} \ (\bd{e}^{\rm obs}-P^{\rm sm}\bd{q}) \  \rangle \ ,
\label{extgamma}
\end{equation}
where $P^{\rm sm}$ is the smear polarisability and
$\bd{q}=\frac{\bd{e^\star}}{P^{\rm sm\star}}$ is the anisotropic
component of the PSF. The symbol $^\star$ indicates those quantities
are measured on stars.  $P_{\gamma}$, defined in \cite {LK97}, is the
correction to the shear polarisability which includes circular
smearing by the PSF.

We compute the stellar quantities, $P^{\rm sh\star}$, $P^{\rm
  sm\star}$ and $\bd{q}$, with the same filter function
  $W\left(\theta,\sigma \right)$ in order to keep the calibration free
  of extra bias. Following \cite{HOEK98} and the STEP results
  \citep{step1,step2}, for each galaxy we compute all quantities,
  including those estimated from the stars, using a filter scale
  $\sigma=r_g$ as given by {\tt IMCAT}.

We did not apply the same PSF anisotropy correction to small and large
objects.  Using the STEP1 and STEP2 simulated catalogues we found that
the measurement of moments from small objects can be significantly
improved and are more robust by first resampling the intensity of
light in each pixel.  Each image is oversampled by a factor of two and
interpolated using a nearest neighbour interpolation kernel, prior to
measure shapes of objects.  The interpolation works very well for
objects with a size close to the star size and does not produce any
detectable extra bias. However, it fails and may even degrade the
signal as the object size increases. STEP simulations show the
transition arises when object size exceeds 1.2$\times$ seeing.

We approximate $P_\gamma$ by half of its trace, ${\rm
Tr}\,{P_\gamma}/2$.  Since individual ${\rm Tr}\,{P_\gamma}$ are
noisy, we derive their values from a fit as function of some galaxy
properties. As described in \cite {step1} and \cite {step2}, the shear
bias parameter $m$ often depends on object sizes, $r_g$, and
magnitudes $mag$.  We therefore fit ${\rm Tr}\,{P_\gamma}$ in the
$r_g$-$mag$ plane using a polynomial that only depends on these two
parameters. The ${\rm Tr}\,{P_\gamma}$-dependence on $mag$ is more
scattered than that on $r_g$, so we choose a function that gives more
weight on $r_g$:
\begin{equation}
  { {\rm Tr}\,{P_\gamma} \over 2} =a_1+a_2 \ r_g + a_3 \ r_g^2 + a_4 \ mag \ .
\end{equation}

\begin{figure*}
  \begin{center}
    \includegraphics[width=6.5cm,height=6cm]{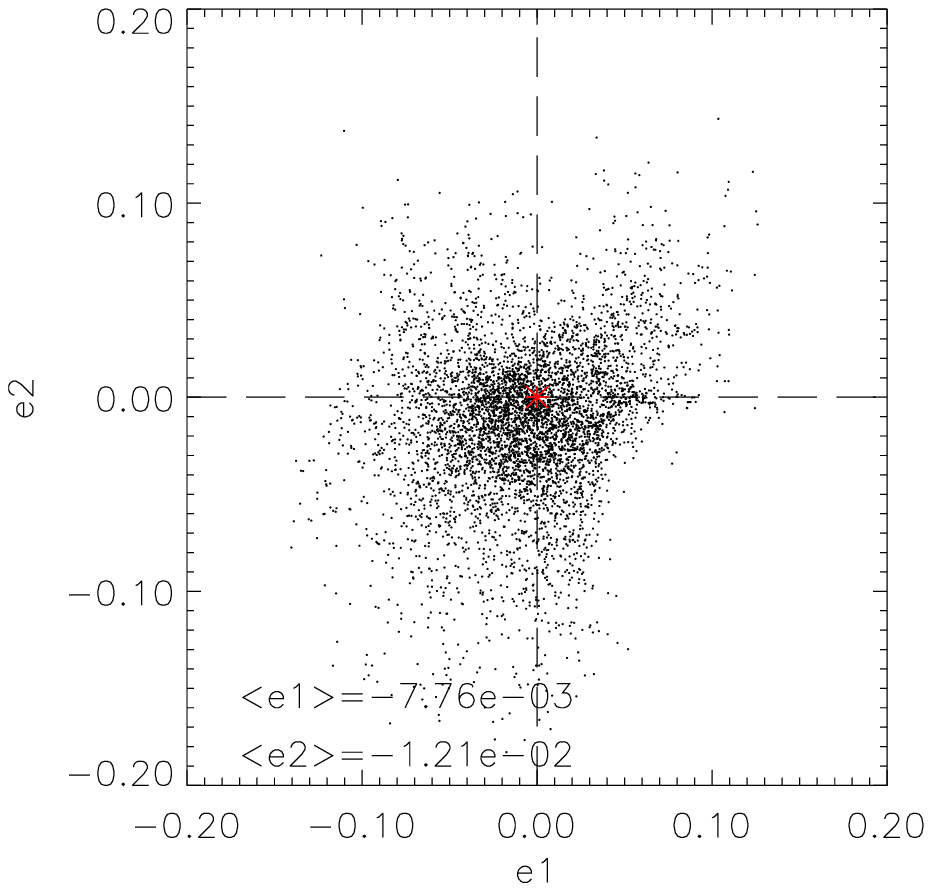}
    \includegraphics[width=6.5cm,height=6cm]{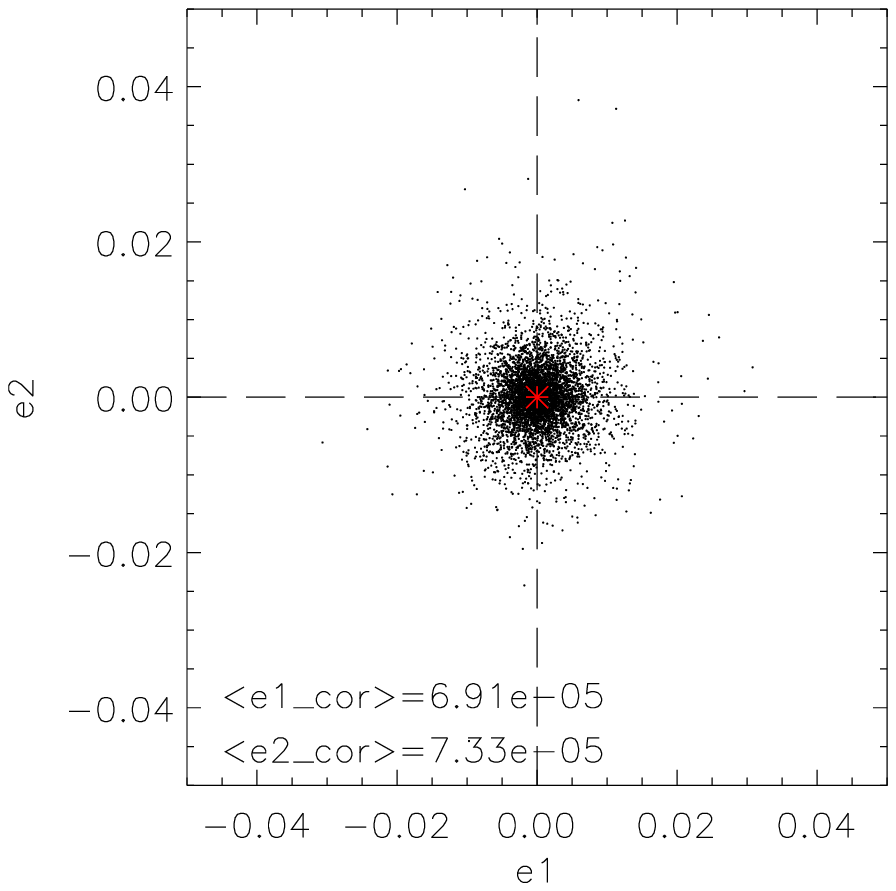}
        \caption{\emph{Left panel:} The observed ellipticities of all
      stars in the 57 pointings. The red cross marks the (0, 0)
      position.  \emph{Right panel:} The residual star ellipticity
      after PSF anisotropy correction. In both plots, the mean values
      of the two ellipticity components are given.}
    \label{psfall}
     \end{center}
\end{figure*}

Table~\ref{pipe} summarises the key parameters of our shear
measurement pipeline.  As an illustration of its application, the left
panel of Fig.~\ref{psfall} shows a compilation of all PSF measurements
for all stars of the 57 pointings used in this work. The distribution
of corrected stars ellipticities in the right panel, shows a reduction
by a factor of 10 in both the average ellipticity and dispersion,
without showing a preferential direction.

\begin{figure*}[!tb]
  \begin{center}
    \includegraphics[width=5cm, height=6.cm]{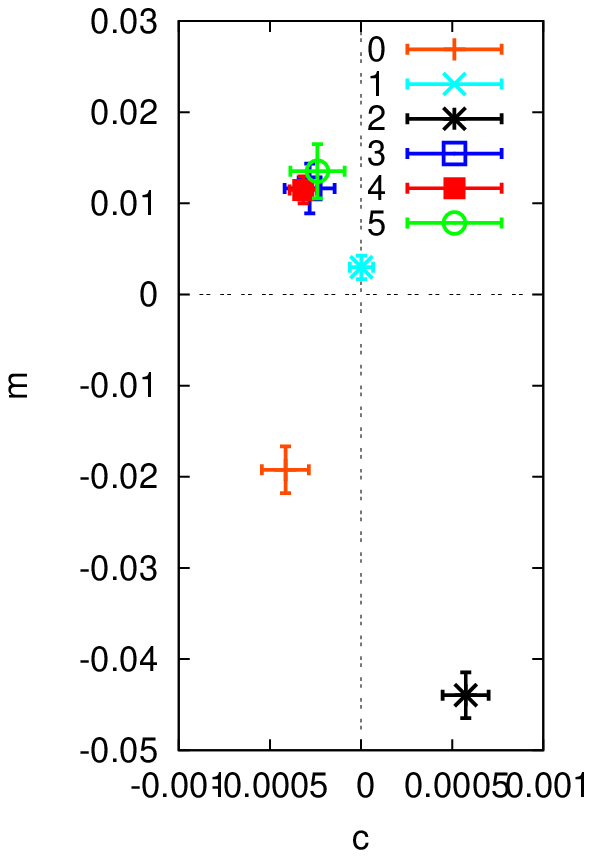}
    \includegraphics[width=5cm, height=6.cm]{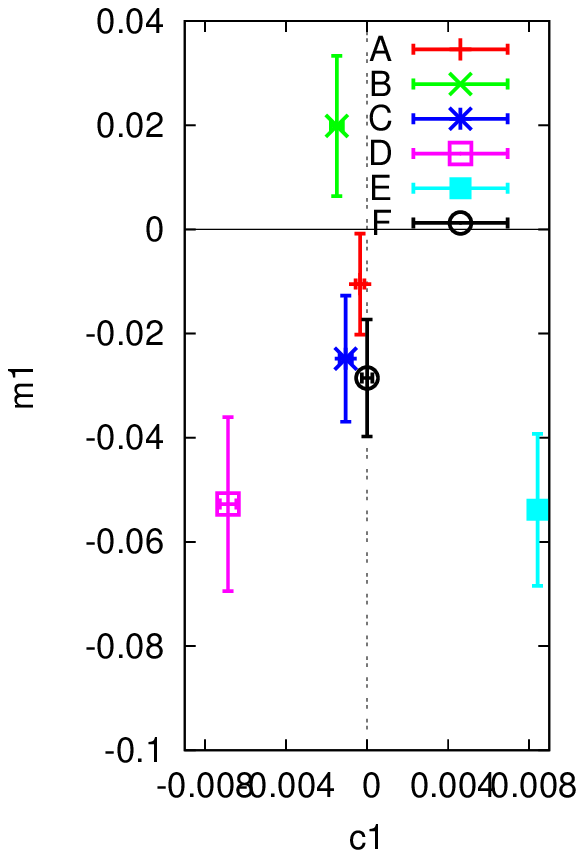}
    \includegraphics[width=5cm, height=6.cm]{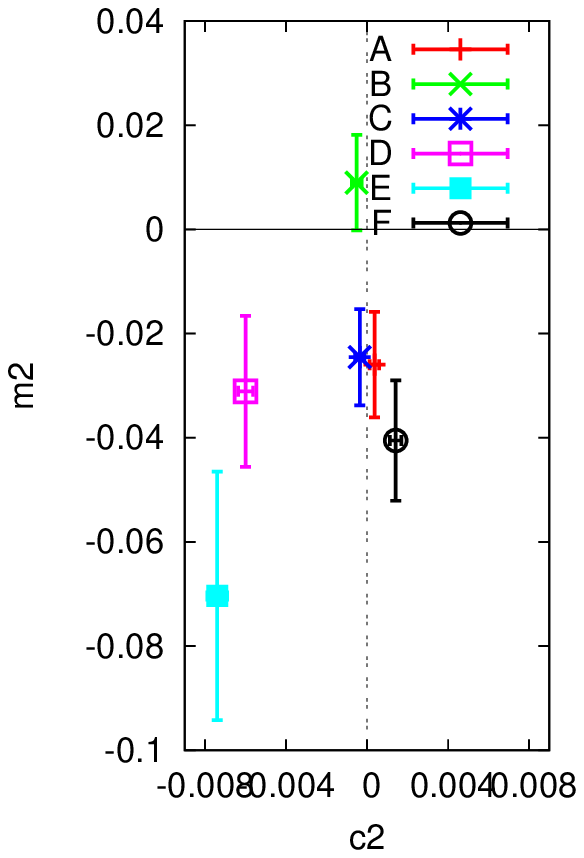}
    \caption{The calibration bias $m$ and the residual offset $c$ of
       our pipeline estimated using STEP simulations. \emph{Left panel:}
      The results of STEP1 for the first component of shear. PSF
      models are labeled from 0 to 5. \emph{Middle and right panels:} The
      results of STEP2 for the two shear components.  PSF models are
      labeled from A to F.}
    \label{step}
  \end{center}
\end{figure*}

Applying the pipeline to the STEP simulations, we quantify the STEP
results using the fit defined in \cite{step1} and \cite {step2}, which
expresses the difference between measured and input shear through a
linear relation:

\begin{equation}
\langle \gamma_i \rangle - \gamma_i^{\rm{input}}=m_i\gamma_i^{\rm{input}}+c_i,
\end{equation}
where $i = 1,2$ are the two shear components.  For a perfect shear
measurement, $m_i$ and $c_i$ would be zero.  Figure~\ref{step} shows the
values of the residual shear offset $c_1$ and of the multiplicative
calibration bias $m_1$ for each of the STEP1 PSF models of the
simulation.  Averaging over the six STEP1 PSF models, our shear
measurement bias is less than $1\%$ as can be seen in Fig.~\ref{step}
(left panel).  The two right panels of Fig.~\ref{step} show the bias
found in the STEP2 simulations, once the rotated and unrotated images
have been merged as described in \cite{step2}.  The two components of
the shear are underestimated by about $3\%$ on average.  Model C is
the most similar to the seeing found in our CFHTLS images.  It is
worth noting that our poorest results come from PSF 2 of STEP1 and
PSFs D and E of STEP2 which have the strongest anisotropy of all the
simulations. Every shear method tested on these particular simulations
had difficulty recovering the correct shear.

These results show the residual bias is well constrained and
reasonably low for our purposes. In fact, the shape measurement bias
is much lower than the total error affecting the cosmological
parameter estimation. It should also be noticed that the STEP1
simulations have characteristics much more similar to the Wide data
than STEP2 simulations. In particular, the structure of the noise and
the PSF types adopted to generate the STEP1 simulations are very close
to the ones of our data.  For this reason the effective bias in the
CFHTLS Wide shear catalogue is expected to be about $1\%$.

\begin{table*}[!bt]
\caption{Summary of the shear measurement pipeline. }
 \label{pipe}
 \begin{center}
 \begin{tabular}{l|l}
 \hline
 \hline
Source Detection & {\em hfindpeaks} \\
PSF: 2D polynomial model & 2nd order fit of $e^\star (r_g)$, $P^{sm\star}(r_g)$  and $P^{sh\star}(r_g)$ \\
Galaxy radius size & $r_g$ from {\em hfindpeaks} \\
Quadrupole,  $P^{sm}$ and $P^{sh}$ estimate & Interpolation (seeing $< r_g < 1.2 \times$ seeing ), $\theta_{\rm {max}}=4r_g$ and  $\Delta \theta = 0.5 \ \rm pixel$ \\
                    & Approx  ( $r_g \geq  1.2 \ \times$ seeing ), $\theta_{max}={\rm Int [}4r_g {\rm ]}$ and  $\Delta \theta = 1 \ \rm pixel$ \\
$P_\gamma$ correction & Fit in $(r_g, mag)$ to $\rm {Tr}(P_\gamma)/2$ \\
Weight & \cite {HH02}\\
Ellipticity cut & $\gamma^2 < 1.0$ \\
Size cut & $r_h > 1.05 \ r_h^\star$ and $1.75 \ {\rm pixel}  < r_g < 6.75 \ \rm pixel$\\
Significance cut & $\nu > 8$\\
$P_\gamma$ cut & $0 < \rm {Tr}P_\gamma/2  < 2$\\
Close pairs & $|d| < 10 \  \rm pixel$ removed\\
\hline
 \hline
\end{tabular}
 \end{center}
 \end{table*}

\section{Shear two-point correlation data}

The data vectors and $1\sigma$ error bars plotted in Fig.~\ref{shear},
for the various shear two-point functions are listed in
Table~\ref{2pttablexi}-\ref{2pttablemap}.

\begin{table*}
\caption{Values of the shear correlation function and the shear
  top-hat variance, as function of scale $\theta$ in arcmin. The
  errors include statistical errors and non-Gaussian-calibrated cosmic
  variance for the E-mode, while only statistical uncertainty contributes to the error of the B-mode.  }
 \label{2pttablexi}
\begin{center}
\newcommand{\rul}{\rule[-2.mm]{0mm}{5.5mm}}
\begin{tabular}{|c|cccc||cccc|}
\hline\hline
\rul $\theta$ & $\xi_{\rm E} $ & $ \xi_{\rm B} $ &$\delta\xi_{\rm E}$ & $\delta\xi_{\rm B}$ &   $\langle|\gamma|^2\rangle_{\rm E}$ &
$\langle|\gamma|^2\rangle_{\rm B}$ &$\delta\langle|\gamma|^2\rangle_{\rm E}$ &$\delta\langle|\gamma|^2\rangle_{\rm B}$\\
\hline
1.00     & 9.704e-05     & 2.494e-05     & 3.416e-05     & 3.200e-05     &       1.158e-04       & -1.809e-06    & 1.749e-05     & 1.036e-05 \\
1.18     & 8.548e-05     & 2.546e-05     & 2.947e-05     & 2.739e-05     &       1.054e-04       & -2.083e-06    & 1.566e-05     & 8.874e-06 \\
1.39     & 6.511e-05     & 7.956e-06     & 2.555e-05     & 2.352e-05     &       9.631e-05       & -1.422e-06    & 1.405e-05     & 7.606e-06 \\
1.64     & 6.160e-05     & -1.138e-05    & 2.209e-05     & 2.012e-05     &       8.777e-05       & -1.570e-07    & 1.261e-05     & 6.521e-06 \\
1.93     & 7.294e-05     & -1.479e-05    & 1.925e-05     & 1.730e-05     &       7.933e-05       & 9.634e-07     & 1.133e-05     & 5.593e-06 \\
2.28     & 6.865e-05     & -2.900e-06    & 1.672e-05     & 1.486e-05     &       7.111e-05       & 1.695e-06     & 1.020e-05     & 4.800e-06 \\
2.69     & 4.578e-05     & 1.275e-05     & 1.456e-05     & 1.277e-05     &       6.373e-05       & 2.020e-06     & 9.174e-06     & 4.122e-06 \\
3.17     & 4.167e-05     & 6.394e-06     & 1.279e-05     & 1.101e-05     &       5.689e-05       & 2.544e-06     & 8.276e-06     & 3.542e-06 \\
3.74     & 2.826e-05     & 2.609e-06     & 1.117e-05     & 9.473e-06     &       5.100e-05       & 2.656e-06     & 7.459e-06     & 3.045e-06 \\
4.41     & 4.490e-05     & -1.043e-06    & 9.840e-06     & 8.201e-06     &       4.615e-05       & 2.538e-06     & 6.735e-06     & 2.617e-06 \\
5.20     & 2.325e-05     & 8.105e-06     & 8.646e-06     & 7.050e-06     &       4.190e-05       & 2.210e-06     & 6.081e-06     & 2.247e-06 \\
6.13     & 2.908e-05     & -9.270e-07    & 7.556e-06     & 5.996e-06     &       3.756e-05       & 1.671e-06     & 5.487e-06     & 1.929e-06 \\
7.22     & 2.979e-05     & 1.723e-06     & 6.634e-06     & 5.099e-06     &       3.346e-05       & 1.241e-06     & 4.962e-06     & 1.656e-06 \\
8.52     & 3.179e-05     & -5.419e-06    & 5.846e-06     & 4.398e-06     &       2.986e-05       & 1.020e-06     & 4.483e-06     & 1.421e-06 \\
10.04    & 1.668e-05     & -4.458e-06    & 5.195e-06     & 3.793e-06     &       2.666e-05       & 7.947e-07     & 4.070e-06     & 1.219e-06 \\
11.84    & 1.687e-05     & 2.365e-06     & 4.647e-06     & 3.257e-06     &       2.393e-05       & 5.714e-07     & 3.711e-06     & 1.044e-06 \\
13.97    & 1.688e-05     & 1.362e-06     & 4.189e-06     & 2.748e-06     &       2.154e-05       & 4.431e-07     & 3.408e-06     & 8.937e-07 \\
16.47    & 1.530e-05     & -1.318e-06    & 3.830e-06     & 2.338e-06     &       1.944e-05       & 2.889e-07     & 3.159e-06     & 7.649e-07 \\
19.42    & 1.579e-05     & -3.319e-06    & 3.530e-06     & 2.004e-06     &       1.751e-05       & 1.432e-07     & 2.942e-06     & 6.548e-07 \\
22.90    & 1.353e-05     & 7.628e-07     & 3.296e-06     & 1.727e-06     &       1.578e-05       & -8.315e-08    & 2.763e-06     & 5.609e-07 \\
27.00    & 1.207e-05     & -5.854e-07    & 3.051e-06     & 1.477e-06     &       1.413e-05       & -3.782e-07    & 2.600e-06     & 4.812e-07 \\
31.84    & 9.731e-06     & -1.522e-06    & 2.850e-06     & 1.274e-06     &       1.266e-05       & -7.709e-07    & 2.453e-06     & 4.135e-07 \\
37.54    & 1.057e-05     & -2.188e-06    & 2.696e-06     & 1.110e-06     &       1.138e-05       & -1.175e-06    & 2.316e-06     & 3.562e-07 \\
44.26    & 6.947e-06     & -2.544e-06    & 2.513e-06     & 9.671e-07     &       1.007e-05       & -1.595e-06    & 2.179e-06     & 3.076e-07 \\
52.19    & 9.153e-06     & -5.789e-06    & 2.363e-06     & 8.417e-07     &       8.614e-06       & -1.932e-06    & 2.051e-06     & 2.664e-07 \\
61.54    & 7.506e-06     & -3.359e-06    & 2.239e-06     & 7.335e-07     &       7.216e-06       & -2.172e-06    & 1.922e-06     & 2.315e-07 \\
72.57    & 4.613e-06     & -4.799e-06    & 2.081e-06     & 6.492e-07     &       5.996e-06       & -2.245e-06    & 1.799e-06     & 2.020e-07 \\
85.57    & 1.110e-06     & -2.967e-06    & 1.977e-06     & 5.706e-07     &       4.980e-06       & -2.150e-06    & 1.680e-06     & 1.774e-07 \\
100.90   & 2.006e-06     & -3.665e-06    & 1.873e-06     & 5.117e-07     &       4.159e-06       & -1.927e-06    & 1.567e-06     & 1.570e-07 \\
118.98   & 2.416e-06     & -2.401e-06    & 1.797e-06     & 4.636e-07     &       3.619e-06       & -1.689e-06    & 1.465e-06     & 1.407e-07 \\
140.29   & 1.982e-06     & -1.174e-06    & 1.743e-06     & 4.272e-07     &       3.203e-06       & -1.482e-06    & 1.373e-06     & 1.284e-07 \\
165.42   & 1.672e-06     & -4.094e-07    & 1.727e-06     & 4.059e-07     &       2.745e-06       & -1.171e-06    & 1.301e-06     & 1.206e-07 \\
195.06   & 3.439e-06     & -9.141e-07    & 1.743e-06     & 3.961e-07     &       2.214e-06       & -7.842e-07    & 1.259e-06     & 1.223e-07 \\
230.00   & 1.780e-06     & -1.420e-06    & 1.876e-06     & 4.081e-07     &       1.615e-06       & -2.919e-07    & 1.333e-06     & 1.686e-07 \\

\hline\hline
\end{tabular}
\end{center}
\end{table*}

\begin{table*}
\caption{Values of the the aperture-mass variance, as function of
  scale $\theta$ in arcmin. }
 \label{2pttablemap}
\begin{center}
\newcommand{\rul}{\rule[-2.mm]{0mm}{5.5mm}}
\begin{tabular}{|c|cccc|}
\hline\hline
\rul $\theta$ & $\langle M_{\rm
  ap}^2\rangle$ & $\langle M_\perp^2\rangle$ & $\delta\langle M_{\rm
  ap}^2\rangle$ & $\delta\langle M_\perp^2\rangle$\\
\hline
1.00     & 1.201e-05     & 1.143e-06     & 5.944e-06     & 5.695e-06 \\
1.18     & 1.402e-05     & -3.377e-06    & 5.138e-06     & 4.875e-06 \\
1.39     & 1.448e-05     & -3.000e-06    & 4.447e-06     & 4.173e-06 \\
1.64     & 1.306e-05     & -8.902e-07    & 3.856e-06     & 3.571e-06 \\
1.93     & 1.107e-05     & -2.830e-07    & 3.353e-06     & 3.059e-06 \\
2.28     & 9.292e-06     & -8.651e-07    & 2.923e-06     & 2.622e-06 \\
2.69     & 8.821e-06     & -1.441e-06    & 2.552e-06     & 2.248e-06 \\
3.17     & 8.556e-06     & -1.258e-06    & 2.238e-06     & 1.930e-06 \\
3.74     & 8.472e-06     & -9.641e-07    & 1.973e-06     & 1.657e-06 \\
4.41     & 8.221e-06     & -9.198e-07    & 1.743e-06     & 1.423e-06 \\
5.20     & 7.676e-06     & -7.686e-07    & 1.542e-06     & 1.222e-06 \\
6.13     & 6.249e-06     & -5.371e-07    & 1.364e-06     & 1.048e-06 \\
7.22     & 5.030e-06     & -1.800e-07    & 1.209e-06     & 8.971e-07 \\
8.52     & 4.609e-06     & 3.606e-07     & 1.079e-06     & 7.693e-07 \\
10.04    & 4.508e-06     & 5.421e-07     & 9.612e-07     & 6.611e-07 \\
11.84    & 4.301e-06     & 4.289e-07     & 8.544e-07     & 5.684e-07 \\
13.97    & 3.976e-06     & 3.853e-07     & 7.688e-07     & 4.874e-07 \\
16.47    & 3.526e-06     & 3.019e-07     & 7.046e-07     & 4.171e-07 \\
19.42    & 3.009e-06     & 2.118e-07     & 6.501e-07     & 3.572e-07 \\
22.90    & 2.600e-06     & 1.249e-07     & 5.954e-07     & 3.066e-07 \\
27.00    & 2.239e-06     & 4.597e-08     & 5.387e-07     & 2.633e-07 \\
31.84    & 1.976e-06     & 2.932e-08     & 4.840e-07     & 2.261e-07 \\
37.54    & 1.877e-06     & 8.679e-08     & 4.359e-07     & 1.944e-07 \\
44.26    & 1.675e-06     & 1.961e-07     & 3.967e-07     & 1.674e-07 \\
52.19    & 1.322e-06     & 3.382e-07     & 3.668e-07     & 1.443e-07 \\
61.54    & 1.147e-06     & 4.524e-07     & 3.454e-07     & 1.245e-07 \\
72.57    & 1.303e-06     & 5.266e-07     & 3.312e-07     & 1.077e-07 \\
85.57    & 1.475e-06     & 5.585e-07     & 3.223e-07     & 9.356e-08 \\
100.90   & 1.511e-06     & 4.750e-07     & 3.172e-07     & 8.174e-08 \\
118.98   & 1.476e-06     & 2.712e-07     & 3.158e-07     & 7.206e-08 \\
140.29   & 1.317e-06     & 3.462e-09     & 3.174e-07     & 6.437e-08 \\
165.42   & 9.472e-07     & -9.410e-08    & 3.220e-07     & 5.879e-08 \\
195.06   & 5.901e-07     & -1.055e-07    & 3.345e-07     & 5.692e-08 \\
230.00   & 3.641e-07     & -7.764e-08    & 3.966e-07     & 7.353e-08 \\
\hline\hline
\end{tabular}
\end{center}
\end{table*}

\end{appendix}

\end{document}